\documentclass[showpacs,amsmath,amssymb,aps,twocolumn,superscriptaddress,prx]{revtex4-2}
\usepackage{mathtools}
\usepackage{algorithm}
\usepackage{algpseudocode}
\usepackage{multirow}
\usepackage[pdftex]{graphicx} \graphicspath{{}}% Include figure files
\usepackage{grffile} % change algorithm for searching thru graphics extensions
\usepackage{comment}
\usepackage{makecell}
\usepackage{braket}
\usepackage{placeins}
\usepackage{diagbox}
\usepackage{bbding}
\usepackage{float}
\usepackage{afterpage}
\usepackage{dcolumn}% Align table columns on decimal point
\usepackage{bm}% bold math
\usepackage[pdftex,colorlinks=true]{hyperref}
\hypersetup{
	colorlinks=true,
	linkcolor=blue,
	urlcolor=cyan,
}

\usepackage{color}
\definecolor{ForestGreen}{RGB}{34, 139, 34}

\newcommand{\affA}{School of Physics, Peking University, 100871 Beijing, China.}
\newcommand{\affB}{Max Planck Institute for the Physics of Complex Systems, N\"othnitzer Str.~38, 01187 Dresden, Germany.}
\newcommand{\affC}{Zhiyuan College, Shanghai Jiao Tong University, 200240, Shanghai China}

\begin{document}

\title{{Floquet-engineered Emergent Massive Nambu–Goldstone Modes}
 }

\author{Yang Hou}
 \affiliation{\affA}

 \author{Zhanpeng Fu}
 \affiliation{\affC}
	\author{Roderich Moessner}
 \affiliation{\affB}
	\author{Marin Bukov}
 \affiliation{\affB}
 	\author{Hongzheng Zhao}
	\email{hzhao@pku.edu.cn}
 \affiliation{\affA}
	\date{\today}

\begin{abstract}
We present a general {framework to implement} massive Nambu–Goldstone quasi-particles in driven many-body systems. {The underlying mechanism leverages an explicit} Lie group structure {imprinted} into an effective Hamiltonian {that governs the dynamics of slow degrees of freedom}; {the resulting emergent continuous symmetry is weakly explicitly broken, giving rise to a massive Nambu-Goldstone mode, with} a {spectral mass} gap scaling linearly with the drive period. We discuss {explicit} and {experimentally implementable} realizations, such as Heisenberg-like spin models {that support} gapped spin-wave excitations. We provide a protocol to certify the existence of the massive Nambu-Goldstone mode from the dynamics of specific observables, and analyse the dispersion spectrum and their lifetime in the presence of weak explicit symmetry breaking. 
\end{abstract}
	\maketitle
\let\oldaddcontentsline\addcontentsline
\renewcommand{\addcontentsline}[3]{}

\textit{Introduction.---}
Collective excitations in many-body systems are a {long-standing yet} everlasting subject. Among others, of particular importance are massless Nambu–Goldstone modes (NGs) appearing {due to spontaneous breaking of a continuous symmetry}~\cite{goldstone1962broken}. They are essential for understanding various intriguing many-body phenomena, e.g., superfluidity and phonons in crystals~\cite{watanabe2020counting}.
Yet, most symmetries {in realistic systems} are indeed {only} approximate. The interplay between explicit symmetry breaking (SB) and spontaneous symmetry breaking (SSB) can generate massive Nambu–Goldstone modes (mNGs)~\cite{nambu1961dynamical,weinberg1972approximate,weinberg1995quantum}, e.g., the $\pi$-meson.

Over the past decades, properties of mNGs have been extensively investigated in {equilibrium}~\cite{uchino2010quasi,watanabe2011number,kapustin2012remarks,watanabe2013massive,nicolis2013more,nicolis2013implications,CoherentMagnonOptics2014,liu2015massive,metlitski2011entanglement,hayata2015dispersion,luitz2015universal,kulchytskyy2015detecting,ohashi2017conformal,leonard2017monitoring,guo2019low,beekman2019introduction,geier2021exciting,alaeian2021limit,delacretaz2022damping,shi2023type,diessel2023generalized,khatua2023pseudo,armas2023approximate,zhang2023quantum,roscilde2023rotor,di2023chiral,song2023dynamical,kiselev2023inducing,surace2023weak,de2024simultaneous,ren2024quasi}, including, e.g., counting rules for their numbers~\cite{watanabe2020counting},  mass~\cite{nicolis2013implications,watanabe2013massive} and damping processes~\cite{delacretaz2022damping,armas2023approximate}. In contrast, little has been known in the case of generic time-dependent systems, which are ubiquitous in nature and experiments.
Perhaps one of the simplest setups are time-periodic (Floquet) systems, which have been recently proposed as a toolbox to realize a zoo of novel non-equilibrium phases of matter~\cite{kitagawa2010topological,martinez2016real,potter2016classification,titum2016anomalous,khemani2016phase,else2016floquet,jangjan2022flqouet,yao2017discrete,oka2019floquet,schweizer2019floquet,else2020long,zhao2021random,PhysRevLett.127.050602,geier2021floquet,dumitrescu2022dynamical,petiziol2022non,zhang2022digital,kalinowski2023non,jin2023fractionalized,sun2023engineering,hossein2025magnon}. Yet, the very existence of mNGs in Floquet systems remains largely unexplored, especially when symmetries are not exact. Here, {we pose the questions of whether} one can exploit Floquet protocols to manipulate various explicit SB processes~\cite{fu2024engineeringhierarchicalsymmetries,jin2024floquet}, stabilize mNGs, and even engineer intriguing non-equilibrium features without a counterpart in static systems.

For many physical systems explicitly fine-tuning SB process is a fundamental challenge as it normally occurs beyond our control, like defects or impurities in materials.
Therefore, we mostly consider engineered quantum platforms that allow for the simulation of time-dependent systems in idealized and controlled settings ~\cite{fukuhara2013microscopic,sun2021realization,mi2022time,kranzl2023observation}.

In this work, we build on the Floquet framework to explore whether or not, and under which conditions, one can implement emergent mNGs; we investigate how to tune their fundamental properties, such as mass and lifetime, even when the drive protocol does not preserve any symmetry.
Realizing such a control scheme is a demanding challenge since: 
(1) The existence of mNGs in equilibrium requires a well-defined SSB ground state w.r.t.~a continuous symmetry; however, generic Floquet protocols break all symmetries, and the concept of SSB and its implications become elusive; 
(2) {How to certify the existence of the mNGs from the experimentally accessible dynamical signatures remains unclear.}
(3) Due to the absence of energy conservation, driven systems eventually heat up to an infinite temperature ensemble, obliterating any non-trivial collective behavior. 

To address these challenges, we construct a family of Floquet protocols that lead to emergent mNGs with a parametrically tunable gap size and lifetime. The key conceptual ingredient is a driving scheme that dynamically exhibits an emergent hierarchical symmetry structure~\cite{fu2024engineeringhierarchicalsymmetries}, together with a Lie group structure that guarantees the appearance of mNGs. 
The construction applies to classical and quantum systems alike, irrespective of the underlying microscopic details.
Due to prethermalization-induced suppression of energy absorption at large drive frequencies (compared to local energy scales)~\cite{bukov2015prethermal,mori2016rigorous,abanin2015exponentially}, emergent mNGs can be identified well before the onset of heating. Moreover, we show that they respond differently to various types of explicit SB perturbations: their mass originates from effective processes preserving a subgroup structure, while their lifetime is determined by even weaker effects that eventually break all symmetries.

As a concrete demonstration, we apply our protocol to a classical many-body spin system where NGs correspond to magnon (or spin-wave) excitations. 
To leading order in the drive period $T$, the dynamics is captured by the ferromagnetic (FM) Heisenberg model with a continuous $O(3)$ symmetry; this symmetry degrades to $O(2)$ by a weak effective magnetic field along the $z$-axis, opening up a magnon excitation gap that scales linearly with $T$. 
We perform large-scale and long-time numerical simulations to verify the existence of mNGs from the non-equilibrium dynamics of observables; in particular, we find a quantitative agreement in the excitation spectra obtained numerically from the dynamical structure factor (DSF) and analytically using spin-wave theory. Similar dynamical signatures also occur in quantum systems, and hence can be used as a practical diagnostic of mNGs in experiments with quantum simulators.

Drive-induced higher-order processes break the $O(2)$ symmetry and destabilize the mNGs, leading to characteristic oscillations in quasi-conserved quantities, e.g., the $x/y/z$ components of the total magnetization. Their amplitudes exhibit distinctive scaling behavior, $A_{x/y}{\propto} T$ and $A_{z}{\propto} T^2$, confirming the hierarchical breakdown of the $O(3)$ symmetry in time. Despite their eventual heat death, these mNGs exhibit a parametrically long lifetime: the linewidth of the DSF scales as $\Gamma{\propto} T^3 $ with a scaling exponent different from the naive Fermi’s Golden rule expectation. We rationalize this scaling by applying a recently developed hydrodynamic theory designed to address the limit of weak explicit SB~\cite{delacretaz2022damping}, as is the case in the high-frequency regime. As $\Gamma$ is much smaller than the gap size, drive-induced mNGs constitute well-defined quasi-particles of the Floquet system in the high-frequency regime.

\textit{Protocol.---}
We start by introducing the driving protocol which involves piece-wise constant Hamiltonian evolution. The Floquet operator reads 
\begin{equation}\label{eq.protocol}
\begin{aligned}
U_F{=}U_0^-U_1^-U_0^{+}U_1^-U_2U_1^+U_0^-U_1^+U_0^{+}U_2,
\end{aligned}
\end{equation}
$U^{\pm}_0 {=} e^{\pm iH_0T'}$, $U_1^{\pm} {=} e^{\pm iH_1T'/2}e^{\pm iH'_1T'/2}$, $U_2 {=} e^{- iH_2T'}$, 
where $T^{\prime}{=}{T}/{10}$ is the evolution time for each step and $T$ is the total drive period. 
% Similar construction has been first proposed in Ref.~\cite{fu2024engineeringhierarchicalsymmetries}, where $\pm$ signs are arranged in a way that different explicit SB processes can be effectively echoed out in a hierarchical manner. 
To understand the resulting dynamics, 
one can define a static effective Hamiltonian $Q$ through the relation $U_F{\equiv} e^{-iQT}$. In the high-frequency regime, one can perturbatively determine
$Q{=}\sum_m Q^{(m)}$, with $Q^{(m)}{\sim}T^m$, and its truncation at a finite order, $Q_{\mathrm{eff}}$, can be used to approximate the stroboscopic time evolution. Using an inverse-frequency expansion we obtain the two lowest  order contributions
\begin{equation}\label{eq.effectiveHall}
    \begin{aligned}
        Q^{(0)}{=}\frac{1}{5}H_2,\ 
        Q^{(1)}{=}{-}\frac{iT}{200}([H_1,H_1^{\prime}]{+}2[H_1+H_1^{\prime},H_2]),\\
\end{aligned}
\end{equation}
and $H_0$ only appears in higher-order terms. At early times, the system evolves into a long-lived prethermal plateau with quasi-conserved energy. This prethermal state can be captured by the Gibbs ensemble $\exp(-\beta Q_{\mathrm{eff}})$~\cite{FleckensteinPrethermalization2021}, with the inverse temperature $\beta$ determined by the initial state energy~\footnote{In this letter, temperature is denoted by $1/\beta$ to avoid confusion with the driving period $T$.}. At this stage, concepts and methods in equilibrium systems remain valid, which we use in the following to predict characteristic properties of mNGs and interpret the collective motion of our
driven system.

{In general, mNGs can be implemented by this protocol if $Q^{(0)}$ has a continuous symmetry. Yet, here we specifically require that $Q^{(0)}$ preserves one non-Abelian symmetry group, $G_2$, such that one can separately tune the mass and lifetime of mNGs by different higher-order processes, as elaborated below.}
For generic drives the first-order correction $Q^{(1)}$ breaks all subgroups of $G_2$. 
However, one can use the following construction to break the symmetry hierarchically.
An illustrative case is when $H_1$ and $H'_1$ are proportional to two non-commuting generators of $G_2$, say $X_i$ and $X_j$. Denoting the non-vanishing structure constants as $c_{ijk}$, with $[X_i,X_j]{=}\sum_k c_{ijk}X_k$, allows us to write $Q^{(1)}$ as a superposition $Q^{(1)}{=}\mu \sum_k c_{ijk}X_k{\equiv} \mu M$ where $\mu$ denotes the chemical potential~\footnote{The coefficients $c_{ijk}$ should contain non-vanishing elements, otherwise the symmetry becomes abelian. Then $Q^{(2)}$ breaks the remaining symmetry and opens the mNG gap. However, due to the absence of the non-abelian group structure, determining this gap becomes complicated and requires a case-by-case study. Furthermore, since $Q^{(2)}$ does not preserve any symmetry, mNGs can be quickly destabilized, see detailed discussions in the~Sec.~SM~4~\cite{SM}.}.

Consequently, $Q^{(0)}+Q^{(1)}$ preserves a subgroup $G_1\subset G_2$, spanned by the generators that commute with $M$. Suppose $|0\rangle$ labels the ground state of $Q^{(0)}{+}Q^{(1)}$ and the generator $M$ is broken (either explicitly or spontaneously). Ref.~\cite{watanabe2013massive} showed that generators that commute with $M$ can generate standard massless NGs, but non-commuting generators necessarily excite mNGs; the corresponding masses, exactly determined by group theory, are proportional to the chemical potential. In this example, $\mu$, and therefore the mass of the emergent mNGs, is linear in $T$. 

Since we do not assume any further symmetry structure in $H_0$, the symmetry group $G_1$ will be degraded to the trivial group by the next-order (in $T$) term $Q^{(2)}$. Therefore, although both explicit SB and SSB are allowed by the effective Hamiltonian $Q^{(0)}{+}Q^{(1)}$, only explicit SB occurs in the exact Floquet protocol. This is of significant importance when seeking quantitative descriptions of mNGs, and two possible scenarios need to be addressed separately: $\ket{0}$ is non-degenerate (Case 1) or degenerate (Case 2), w.r.t. the symmetry $G_1$. 

In Case 1, $\ket{0}$ is unique and independent of $T$, and $Q^{(2)}$ is a negligible perturbation. Hence, $Q^{(0)}{+}Q^{(1)}$ is sufficient to predict the masses and dispersion of the mNGs {to a good approximation}. 
Case 2 is more subtle since $Q^{(2)}$ breaks $G_1$ and lifts the degeneracy; hence, one has to take at least $Q^{(2)}$ into account to determine the new ground state~\footnote{In general, the new ground state will explicitly depend on $T$ and the concrete form of $Q^{(2)}$, which contains complicated nested commutators of each Hamiltonian in the drive, making the analytical prediction of the mNGs very difficult in practice. We also note that the inverse-frequency expansion for Floquet systems does not converge in general. However, as a dynamic consequence of this divergence, Floquet heating can be delayed up to exponentially long times. We emphasize that this is irrelevant to the Goldstone physics discussed here, which only persists for algebraically long times.}. For simplicity, in the following, we focus on Case 1 and briefly comment on Case 2 in the end; additional discussions are presented in the Supplementary Material (SM)~\cite{SM}.

Higher corrections of order $\mathcal{O}(T^{2})$ also lead to a finite lifetime of mNGs at any finite drive period $T$; however, their influence can be systematically pushed to later times by increasing the driving frequency.

The proposed construction is generic as it relies only on the aforementioned group structure; in particular, it is independent of the microscopic details of the drive, and hence can be applied to various physical systems of interest. Indeed, we emphasize that there exist many other protocols to Floquet engineer mNGs with similar properties. In the~Sec.~SM~1.1~\cite{SM}, we provide a comprehensive overview of a family of protocols (some of them involving fewer steps than Eq.~\eqref{eq.protocol}), 
which can be efficiently implemented, e.g., on gate-based digital quantum simulators~\cite{geier2021floquet,mi2022time,zhao2023making}; in Sec.~SM~1.2~\cite{SM} we also present a continuous drive feasible for analog quantum platforms~\cite{xu2018emulating,sun2021realization,guo2024site,braun2024real}. This variety of different protocols
also highlights that our proposed Floquet-engineered mNGs is a fundamental and fairly universal emergent
phenomenon.

\textit{Effective Description.---}
This construction applies equally to quantum and classical systems. In the latter, the derivation of the effective Hamiltonian can be obtained by replacing the commutator $-i[\cdot, \cdot]$ in Eq.~\eqref{eq.effectiveHall} with the Poisson bracket $\{\cdot,\cdot\}$~\cite{mori2018floquet}. Hence, to verify the existence and properties of mNGs, especially the ones at a long wavelength, we focus on a 1D classical spin chain with spin-wave excitations, for which the long-time and large-size numerical simulation can be efficiently implemented~\cite{howell2019asymptotic,ye2021floquet,pizzi2021classical,jin2022prethermal,mcroberts2023prethermalization}.  
We consider the Hamiltonians 
\begin{equation}\label{eq.fieldH}
    \begin{aligned}
         H_2&=-J_2\sum\nolimits_{j}\boldsymbol{S}_j\cdot\boldsymbol{S}_{j+1},\ \  H_1=-h_x\sum\nolimits_{j}S_j^x, \\ 
         H_1^{\prime}&=-h_y\sum\nolimits_{j}S_j^y, \ \ H_0=J_x\sum\nolimits_{j}S_j^xS_{j+1}^x,
    \end{aligned}
\end{equation}
 with $N$ sites and periodic boundary conditions.
 The normalized classical spin variable $\boldsymbol{S}_j$ on site $j$ can also be parametrized by azimuthal angle $\phi_j$ and polar angle $\theta_j$ as $S_j^x{=}\sin{\theta_j}\cos{\phi_j}$, $S_j^y{=}\sin{\theta_j}\sin{\phi_j}$, $S_j^z{=}\cos{\theta_j}$; it satisfies the Poisson bracket $\{S_i^\mu,S_j^\nu\}{=}\delta_{ij}\epsilon^{\mu\nu\rho}S_j^{\rho}$ with the fully antisymmetric tensor $\epsilon^{\mu\nu\rho}$. Here $H_2$ corresponds to the Heisenberg model, and $H_1,H_1'$ are generators of $O(3)$ symmetries;
 \begin{figure}[t!] \includegraphics[width=0.883\linewidth]{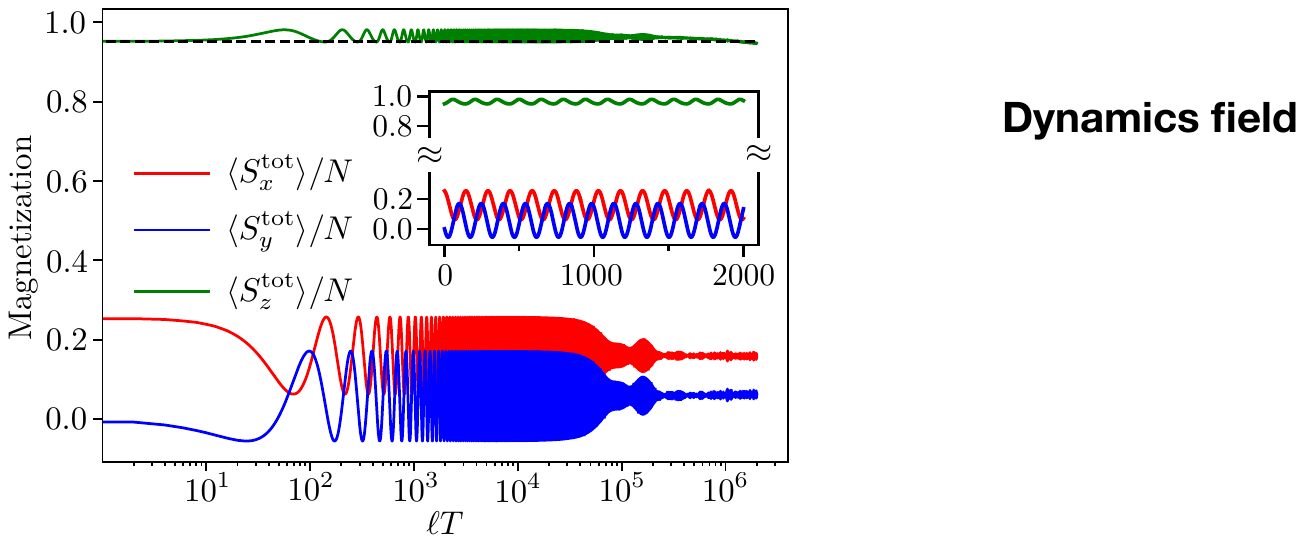}
		\caption{Dynamics of the total magnetizations in a 1D driven Heisenberg-type model. Magnetization in the $(x,y)$-plane (red and blue) rotates periodically, suggesting the existence of mNGs. $\langle S_z^\text{tot}\rangle/N$ (green) oscillates on top of the conserved value (dashed line) due to higher order effects that break $O(2)$ symmetry. We use parameters $T{=}2$, $J_2{=}1$, $h_x{=}2$, $h_y{=}{-}2$, $J_x{=}2$, $\theta_j{=}0.2\pi$, and $\phi_j$ randomly sampled within $(-\pi/3, \pi/3)$.
}
\label{fig.dynamicsfield}
\end{figure} 
 $h_{x/y}$ are the magnetic field strength along the $x/y$-directions, and $J_x$ denotes the interaction strength along $x$-direction which breaks all continuous symmetries. The spin dynamics is described by $\dot{S}_j^\mu {=} \{S_j^\mu,H(t)\}$, where the time-dependence in the Hamiltonian follows the piecewise protocol, Eq.~\eqref{eq.protocol}.
 Following Eq.~\eqref{eq.effectiveHall} we obtain the effective Hamiltonian
\begin{equation}
\label{eq.truancatedH}
Q_{\text{eff}}=-\frac{J_2}{5}\sum\nolimits_{j}\boldsymbol{S}_j\cdot\boldsymbol{S}_{j+1}+\frac{h_xh_yT}{200}\sum\nolimits_{j}S_j^z,
\end{equation} up to order $\mathcal{O}(T)$. 
$Q^{(0)}$ is the Heisenberg model preserving $O(3)$ and $Q^{(1)}$ degrades the symmetry to $O(2)$ around the $z$-axis. For $J_2>0$, the ground state features FM order (Case 1) along the positive (negative) $z$-direction if we take $h_x$ and $h_y$ of the opposite (same) sign~\footnote{Note, Mermin-Wagner theorem does not apply here as the $O(3)$ symmetry is explicitly broken by a $z$-field, which can lead to finite magnetization even at a finite temperature.}.

In the high-frequency regime, the early-time spin dynamics approximately follows the EOM generated by $Q_{\text{eff}}$. If $h_x$ and $h_y$ have the opposite sign and 
considering the regime $S_j^z{\approx} 1$ and $S_j^{x/y}{\ll} 1$, one obtains the spin-wave spectrum via a standard linearization method
\begin{equation}\label{eq.spf}
    \omega=\pm\left[\frac{2J_2}{5}(1-\cos{q})+\frac{|h_xh_y|T}{200}\right],
\end{equation}
with $q$ denoting the quasi-momentum, cf.~Sec.~SM~2.1~\cite{SM} for details.

Similar to the Heisenberg model, a quadratic dispersion appears for long-wavelength modes ($q{\rightarrow} 0$). Yet, at $q{=}0$ this mode has a gap of size $\left|{h_xh_yT}\right|/200$ that is linear in $T$, matching our generic expectation as elaborated above. The corresponding spin dynamics features a total magnetization in the $(x,y)$-plane rotating around the $z$-axis, with a frequency exactly matching the gap size. This dynamical behavior together with the dispersion relation thus provide a distinctive signature of the existence of mNGs that will be numerically verified below.

\begin{figure}[t!] \includegraphics[width=\linewidth]{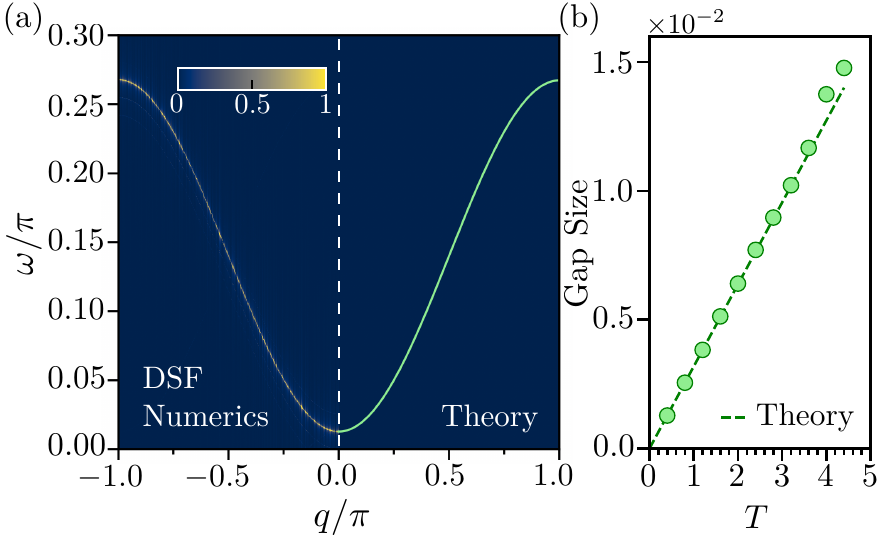}
		\caption{(a) Dynamical structure factor (DSF) through the first Brillouin zone. The left half displays numerical results, which precisely follow the theoretical prediction (right half) and certify the existence of mNGs. We use the drive period $T{=}2$. The Fourier transform is performed by using $10^4$ Floquet cycles. (b) A gap is opened at $q{=}0$ with a tunable size that depends linearly on $T$. The gap size deviates from the theoretical prediction at large $T$. We use $J_2{=}1$, $h_x{=}2$, $h_y{=}{-}2$, $J_x{=}1$, $\theta_j{=}0.01\pi$, and $\phi_j$ randomly sampled within {$(-\pi/3, \pi/3)$.}
}
\label{fig.spectrumfield}
\end{figure}

\textit{Numerical Results.---}
We now confirm the mNGs and discuss various hierarchical SB effects by investigating the Floquet spin dynamics. Initially, the system is prepared as an ensemble where spins deviate from the $z$-axis by a small angle $\theta_j$, 
yielding a sufficiently large z-magnetization density, as required by the spin-wave approximation. The azimuthal angle $\phi_j$ is randomly sampled within a range s.t. the spatial randomness is sufficiently strong to generate many-body effects, but keeps a non-vanishing total magnetization in the $(x,y)$-plane. 

For each driving step, the time evolution is governed by the EOM generated by the corresponding static Hamiltonian, except for
$H_2$ where a symmetric Trotter decomposition is used for numerical efficiency~\footnote{Except for those generated by $H_2$, the EOMs can be explicitly integrated over the corresponding step duration.
For $H_2$, we use a symmetric Trotter decomposition to approximate the evolution~\cite{mcroberts2023prethermalization}, thereby significantly improving the efficiency of the numerical simulations, cf.~details in Sec.~SM~5~\cite{SM}. This method introduces simulation errors at order $\mathcal{O}(T^2)$, and hence it neither changes the form of $Q_{\mathrm{eff}}$, Eq.~\eqref{eq.truancatedH}, nor gap size predicted above.}.
For all numerical results we use $N{=}1000$ and perform ensemble averages over $100$ realizations. 
The dynamics of the total magnetization components is shown in Fig.~\ref{fig.dynamicsfield}. 
The rotating magnetization in the $(x,y)$-plane (red and blue) is a characteristic signature of the mNGs.
The oscillation is periodic in time (clearly visible in the inset where a linear time scale is used) and persists for a controllably long time without exhibiting any notable decay up to $\ell T{\approx}10^5.$ Although $Q_{\text{eff}}$ in Eq.~\eqref{eq.truancatedH} conserves the $z$-magnetization and predicts a static $\langle S_z^\text{tot}\rangle/N$ (dashed line), the exact Floquet protocol indeed generates an oscillatory dynamics (green) since $Q^{(2)}$ breaks the $O(2)$ symmetry. Oscillations in the $(x,y)$-plane gradually damp out and settle into a prethermal plateau, which eventually heat up to an infinite temperature ensemble, cf.Sec.~SM~2.7~\cite{SM}. 
\begin{figure}[t!] \includegraphics[width=0.75\linewidth]{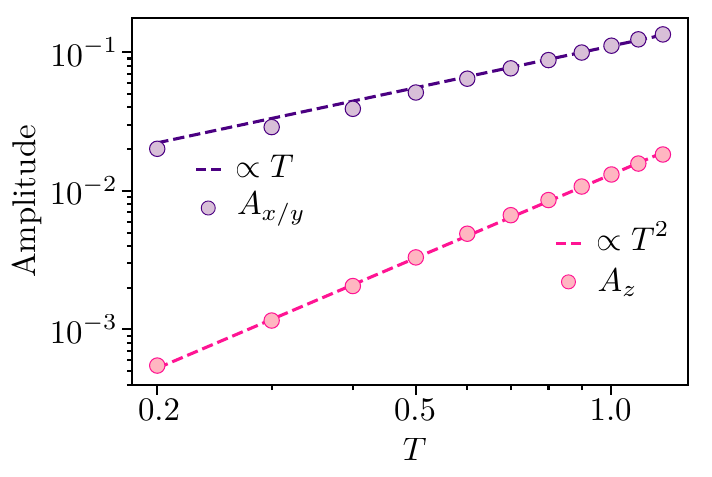}
		\caption{Scaling of oscillation amplitudes, $A_{x/y}{\propto} T$ and $A_z{\propto} T^2$. It suggests that explicit symmetry breaking occurs hierarchically. The same parameters are used as in Fig.~\ref{fig.spectrumfield}.
}
\label{fig.Afield}
\end{figure}

We further verify the entire mNG spectrum by analyzing the DSF $\mathcal{S}(q,\omega){=}\sum_{\alpha{=}x,y}\langle|{S}_{q}^\alpha (\omega)|\rangle$, where ${S}_{q}^\alpha  (\omega)$ is the space-time Fourier transform of the spin configurations, {${S}_{q}^\alpha (\omega){=}\sum_j \int \text{d}t {S}_{j}^\alpha(t)e^{-i(\omega t-qj)}$}. 
Fig.~\ref{fig.spectrumfield}(a) depicts  $\mathcal{S}(q,\omega)$ for a fixed $T$: the spectrum precisely follows our theoretical prediction (green line in the right half), Eq.~\eqref{eq.spf}. A mNG gap opens at $q{=}0$
corresponding to the oscillatory frequency of the total magnetization. It is worth highlighting the controllability of the gap size by tuning the drive period, which also exhibits quantitative agreement with the theoretical prediction (green dashed line) over a wide range of $T$. A noticeable deviation appears when $T$ becomes large, $T\geq 3.5$, as shown in Fig.~\ref{fig.spectrumfield}(b). Note, a faint gapless line also appears in Fig.~\ref{fig.spectrumfield}(a), which is generated by non-linear and drive-induced higher-order effects, see details in Sec.~SM~2.3~\cite{SM}.

$Q^{(2)}$ that is not captured by Eq.~\eqref{eq.truancatedH} breaks all symmetries, and hence modifies and destabilizes the mNGs. The most notable phenomenon is the oscillatory dynamics in $\langle S_z^\text{tot}\rangle/N$ 
as seen before in Fig.~\ref{fig.dynamicsfield}: it occurs because higher order corrections can effectively generate extra fields in the $(x,y)$-plane, slightly tilting the original explicit SB generator of the $O(2)$ symmetry in $Q_{\mathrm{eff}}$. For low-temperature initial states and at short times, the three spin directions are indeed all coupled together by this field; therefore, they share the same oscillating frequency, cf.~inset of Fig.~\ref{fig.dynamicsfield}. However, one clearly observes a difference in the oscillating amplitudes, with $A_{x/y}$ significantly larger than $A_{z}$. More precisely, in Fig.~\ref{fig.Afield} we confirm their asymptotic scaling dependence on $T$, $A_{x/y}{\sim} T$ and ${A_z\sim} T^2$, which can also be analytically justified by a perturbative argument, cf.~Sec.~SM~2.2~\cite{SM}. 

\begin{figure}[t!] \includegraphics[width=\linewidth]{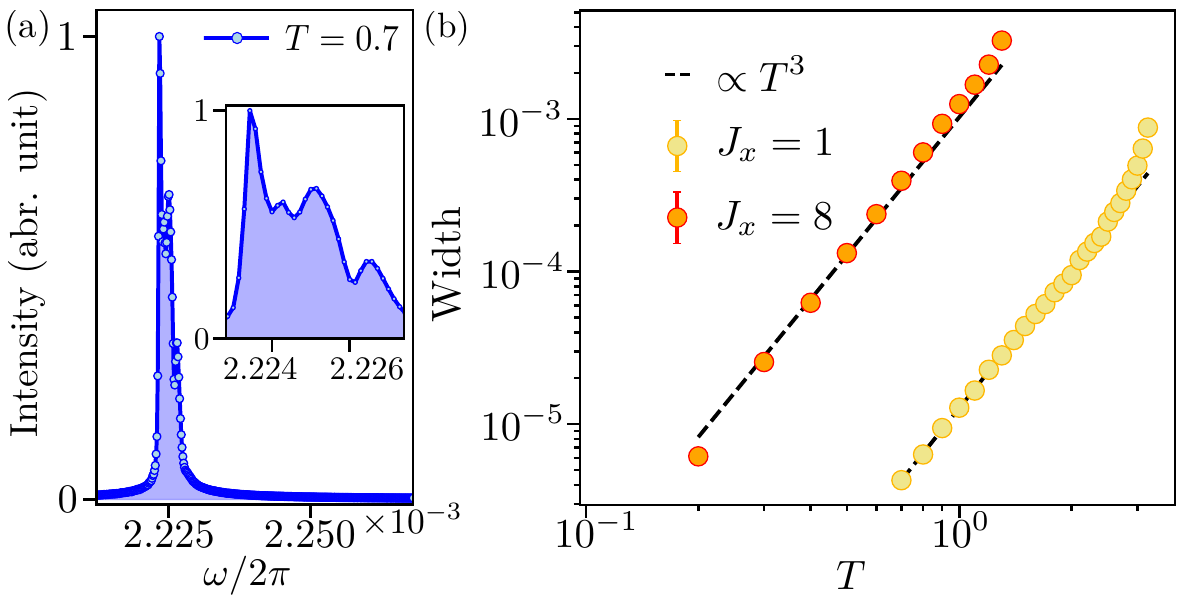}
		\caption{(a) Fourier density for $J_x{=}1$. Damping of mNGs broadens the spectrum, leading to asymmetric lineshape and multiple side peaks that are shown in the inset. (b) Linewidth $\Gamma$  exhibits a power-law scaling in the high frequency regime; the black dashed line with scaling $T^{3}$ is a guide to the eye. This scaling is different from the Fermi’s Golden Rule expectation and can be justified by a hydrodynamic theory. We use $J_2{=}1$, $h_x{=}2$, $h_y{=}{-}2$, $\theta_j{=}0.25\pi$ and $\phi_j$ randomly sampled within $(-\pi/2, \pi/2)$.
}
\label{fig.widthfield}
\end{figure}

For even longer times, higher order effects together with thermal fluctuations (set by the spatial randomness in the initial states) lead to the damping of mNGs which broadens the spectrum. In Fig.~\ref{fig.widthfield} (a) {we depict the Fourier spectrum at $q{=}0$. Strikingly, we observe an asymmetric lineshape and multiple side peaks of comparable heights, one intriguing non-equilibrium
feature that normally does not exist in systems without driving. They depend on the details of the Trotterization protocol being used and also on the specific form of the SB processes, cf.~Sec.~SM~2.4~\cite{SM}.}

The overall linewidth $\Gamma$ in the DSF can be used to characterize the damping rate. It is determined by averaging the full width at $10\%$, $15\%$, and $20\%$ of the maximum intensity of the Fourier peak, with their standard deviation serving as the error bar. $\Gamma$ exhibits a power-law scaling $T^{\alpha}$, suggesting that the mNGs lifetime can be systematically prolonged by increasing the drive frequency. Since all symmetries are broken by $\mathcal{O}(T^2)$ processes, following Fermi's Golden rule one may expect a scaling exponent $\alpha{=}4$~\cite{ikeda2021fermi,oeda2024prethermal,mori2022heating}; however, surprisingly, we find  $\alpha {\approx} 3$, significantly deviating from this naive expectation.

We provide one possible explanation following a recent work~\cite{delacretaz2022damping}, where the authors use hydrodynamics to analyse the damping rate of mNGs at the long wavelength limit.
In the absence of explicit SB effects, for small $q$ the magnons disperse as
$
    \omega {=} {\pm} fq^2{-}i\gamma q^4{+}\cdots,
$
where $f$ determines the dispersion for the isotropic FM Heisenberg model; $\gamma$ is generally non-zero at a finite temperature, originating from magnon-magnon scattering and thermal fluctuations~\cite{tahir1962use,harris1968energy,halperin1969hydrodynamic}.
When symmetries are weakly broken, this dispersion modifies to $
      \omega {=} \pm f(q_0^2+q^2){-}i(\Gamma+D_0q^2{+}\gamma q^4){+}\cdots,
$ with three additional parameters: the damping rate $\Gamma$, attenuation $D_0$ and the gap $q_0^2$. Locality of constitutive relations in hydrodynamics implies that the damping rate for $q{\to} 0$ is completely
determined by the relation $\Gamma=q_0^2(D_0{-}\gamma q_0^2)$~\cite{delacretaz2022damping}.

Supposing the symmetry is broken only by a $z$-field as in  Eq.~\eqref{eq.truancatedH}, mNGs are stable for sufficiently low temperature, hence $D_0{=}0$. Therefore, contributions to $D_0$ should only involve $Q^{(2)}$ and higher-order perturbations. Together with the fact that the gap depends linearly on $T$, we know $q_0^2{\sim} \mathcal{O}(T)$ and arrive at
    $\Gamma {\sim} \gamma\mathcal{O}(T^2){+}\mathcal{O}(T^3)$.
In all numerical simulations performed here, we consider initial states with a large $z$-magnetization density and small thermal fluctuations, hence $\gamma$ can be extremely small. Therefore, the leading order in $\Gamma$ can start from $\mathcal{O}(T^3)$. This indeed matches well with the numerical results in Fig.~\ref{fig.widthfield} (b), especially when the driving segment $H_0$ that breaks the $O(2)$ symmetry is weak, $J_x{=}1$.

\textit{Discussion.---}
The scaling exponent of the mNGs lifetime may not be universal and can depend on Hamiltonian parameters. For instance, for the orange data shown in Fig.~\ref{fig.widthfield}(b) with substantially stronger $O(2)$ SB effects, $\alpha$ is indeed larger than 3, see more examples in Sec.~SM~2.6~\cite{SM}. To explain this scaling behavior one needs to further analyse the scattering channels between mNGs modes, e.g. the two- or three-magnon scattering matrix. Given different SB processes and thermal fluctuations, the dominant effects in $D_0$ may scale differently with $T$, and will be systematically studied in future work. As the scattering matrix now becomes tunable via Floquet-engineering, our work paves a new pathway to stabilize or create novel magnon bound states ~\cite{fukuhara2013microscopic,macri2021bound,kranzl2023observation,turner2024stable,Realization2024}, as well as different transport behaviors which may not exist in non-driven systems. Furthermore, for other values of $q$, $\Gamma$ scales differently with $T$; in particular, $\Gamma$ becomes almost independent of $T$ for large $q$, indicating that thermal fluctuations in the initial state and $Q^{(0)}$ dominate the damping, see details in Sec.~SM~2.5~\cite{SM}.

While we have discussed the emergent mNGs in a classical Floquet system, we emphasize that our Floquet protocol applies equally to both quantum and classical systems. In Sec.~SM~2.9~\cite{SM}, we present numerical simulations of a quantum system and certify the existence of mNGs~\cite{marinquspin},
illustrating the possibility for experimentally detecting the mNGs in real quantum simulators.

Going beyond the present systems, it is intriguing to apply our general construction to physical systems exhibiting different types of mNGs.  
For instance, one can consider the Heisenberg model with
anti-ferromagnetic couplings: a weak external $z$-field leads to a ground state manifold which preserves a $O(2)$ symmetry but this degeneracy is then lifted by $Q^{(2)}$ in driven systems (Case 2). As illustrated in Sec.~SM~3~\cite{SM}, one mNG can still be detected from the spin dynamics. However, we observe strong evidence that the damping effect is substantially different from the FM case, opening up new future directions~\footnote{Indeed, $Q^{(0)}+Q^{(1)}$ also predicts a gapless mode with linear dispersion. However, it is unstable against higher order perturbations and we do not observe them in the exact Floquet dynamics.}.  

Furthermore, the requirement that $Q^{(1)}$ forms a superposition of generators of the non-Abelian group $G_2$, is not necessary for the appearance of mNGs, so long as $Q^{(1)}$ still preserves a subsymmetry of $G_2$.
We give one explicit example in the Sec.~SM~2.8~\cite{SM} where $Q^{(1)}$ involves three-body interactions, which effectively generates a $z$-field for sufficiently small thermal fluctuations in the initial state. {Yet, unlike the model in Eq.~\eqref{eq.truancatedH} where the gap size can be exactly determined,} here the gap can only be computed approximately. Also, this mode tends to be very fragile with a notable broadening of the spectrum linewidth that scales linearly in $T$, see details in Sec.~SM~2.8~\cite{SM}. 

Generalizing the discussion to systems with long-range interactions~\cite{song2023dynamical} and higher-form symmetries~\cite{sala2020ergodicity,yuan2020fractonic,hidaka2021counting,armas2024approximate,gromov2024colloquium,boesl2024deconfinement}, where anomalous or fractonic NGs may exist, also remains interesting to explore. Finally, beyond the Floquet paradigm, we envision the occurrence of mNGs in quasi-periodically~\cite{PhysRevLett.120.070602,zhao2019floquet,else2020long,long2024topologicalphasesmanybodylocalized} and randomly driven systems~\cite{moon2024experimentalobservationtimerondeau}.

\textit{Acknowledgments.---}%
This work is supported by the National Natural Science Foundation of China (Grant No. 12474214),
and by Innovation Program for Quantum Science and Technology (No. 2024ZD0301800), and by Beijing Natural Science Foundation (Grant No.~QY24016) and by the Deutsche Forschungsgemeinschaft under the cluster of excellence ct.qmat (EXC 2147, project-id 390858490), and by “The Fundamental Research Funds for the Central Universities, Peking University”, and by "High-performance Computing Platform of Peking University". 
This work is funded by the European Union (ERC, QuSimCtrl, 101113633). Views and opinions expressed are however those of the authors only and do not necessarily reflect those of the European Union or the European Research Council Executive Agency. Neither the European Union nor the granting authority can be held responsible for them. 
        
\bibliography{sample}

\clearpage

 \let\addcontentsline\oldaddcontentsline
	\cleardoublepage
	\onecolumngrid
 \begin{center}
\textbf{\large{\textit{Supplementary Material} \\ \smallskip
	Floquet-engineered Emergent Massive Nambu-Goldstone Modes}}\\
		\hfill \break
		\smallskip
	\end{center}
	\renewcommand{\thefigure}{S\arabic{figure}}
        \setcounter{figure}{0}
    \renewcommand{\thesection}{SM\;\arabic{section}}
	\setcounter{section}{0}
	\renewcommand{\theequation}{S.\arabic{equation}}
        \setcounter{equation}{0}
    \renewcommand{\thesubsection}{\arabic{subsection}}
	\setcounter{section}{0}
    \tableofcontents
    \setcounter{page}{1}

\section{Alternative Drive Protocols}
\subsection{Discrete Driving Protocol}
\begin{table*}[b]
\caption{Different Protocols for Floquet-engineered mNGs}
    \label{tab:1}
    \centering
    \tabcolsep=0.01\linewidth
    \begin{tabular*}{\linewidth}{c | c c c c} \hline \hline
          Protocol&   \thead{Absence of Symmetry \\in Instantaneous Hamiltonians}&\thead{Hierarchical Symmetry Structure\\in Effective Hamiltonian }&{ Generator Protection in $\mathcal{O}(T)$}&Driving Steps
    \\\hline
           \romannumeral1& ${\times}$ & \checkmark &\checkmark &  14\\
          \romannumeral2& $\times$ &$\times$& $-$ &  4\\
 \romannumeral3& \checkmark &\checkmark &$\times$  & 4 \\
 \romannumeral4 &\checkmark &\checkmark &\checkmark & 8\\\hline\hline
    \end{tabular*}
    \label{tab.protocols}
\end{table*}

In the main text, we propose a protocol that gives rise to mNGs with tunable mass and lifetimes in driven many-body systems. However, these features arise also in other time-dependent protocols. In this section, we discuss four versatile protocols with respective advantages and shortcomings concerning the experimental accessibility and the stability of mNGs, as summarized in Table~\ref{tab.protocols}.
More precisely, we focus on the following features:

\begin{itemize}
    \item[(1)] \textbf{No requirement of perfect symmetry in the instantaneous Hamiltonian generators:}  Modern quantum simulators do not realize perfectly symmetric Hamiltonians. Therefore, in practice, designing a Floquet scheme that does not require the implementation of perfect symmetries, is of great importance to Floquet engineering mNGs. Furthermore, at the fundamental level, if the instantaneous Hamiltonians do not preserve any symmetry, they should not host any mNGs. Therefore,    
    the appearance of Floquet-engineered mNGs after the entire driving cycle, becomes a truly emergent non-equilibrium phenomenon, which does not exist without the drive.

    \item[(2)] \textbf{Hierarchical symmetry structure in effective Hamiltonian:} This implies that, $Q^{(0)}$ preserves a non-abelian symmetry group $G_2$, $Q^{(1)}$ preserves a sub-group, $G_1\subset G_2$, and higher-order terms break all remaining symmetries. Even though the entire Floquet operator does not preserve any symmetry, 
    this mechanism ensures the existence of mNGs in the high-frequency limit, and its mass is generated by $Q^{(1)}$ at the order $\mathcal{O}(T)$. However, the determination of the mass of mNGs can be complicated and strongly depends on the details of the microscopic Hamiltonian, see discussions in~\ref{SM.3b}.

    \item[(3)] \textbf{Generator protection of the mass of mNGs in the effective Hamiltonian at the order $\mathcal{O}(T)$:} We can go beyond the aforementioned hierarchical symmetry breaking construction and notice that, if $Q^{(1)}$ precisely corresponds to one of the symmetry generators of $G_2$, mNGs' mass can be determined analytically by exploiting the non-abelian group structure. This feature is particularly important when the usual numerical analysis of mNGs faces difficulties, especially in the thermodynamic limit. It also allows for the exact controllability of the mass magnitude by the Floquet drive.

    \item[(4)] \textbf{Number of driving steps:} The length of a driving protocol characterizes its complexity and short protocols are generally more accessible in experiments. For instance, in gate-based digital quantum platforms, more steps imply that more quantum gates are involved, which inevitably introduces extra gate errors that further destabilize the mNGs.
\end{itemize}

The detailed properties of the four protocols can be summarized as follows:

\textit{Protocol} \romannumeral1.
This is the protocol from Eq.~\eqref{eq.protocol} in the main text. This 14-step piece-wise protocol requires symmetry in the instantaneous Hamiltonians. Despite its complexity, its primary advantage lies in the structure of the effective Hamiltonian, where $Q^{(1)}$ corresponds to the symmetry generator of $G_2$. 
For sufficiently low-temperature initial states, the damping process of mNGs arises from higher-order contributions to the effective Hamiltonian $Q^{(2)}$. Additionally, the EOM generated by the instantaneous Hamiltonian in this protocol is either analytically integrable or can be conveniently decomposed into integrable parts using Trotter decomposition, which significantly facilitates the efficiency of numerical simulations, see details in~\ref{sec.Trotter}.

\textit{Protocol} \romannumeral2.
One can try to simplify the protocol above by considering the 4-step drive
\begin{equation}
U_F=U_0^{+}U_2U_MU_0^{-}=e^{-iTH_0}e^{-iH_2T}e^{-iH_M\theta T}e^{iH_0T},
\end{equation}
where $H_2$ preserves the $G_2$ symmetry, $H_M$ is a generator of $G_2$ and preserves $G_1\subset G_2$, and $H_0$ breaks all symmetries. $U_M$ is a global rotation around the generator $H_M$ for a time-window $\theta T$. Thus, the leading order of the effective Hamiltonian reads
\begin{equation}
    Q^{(0)}=H_2+\theta H_M,
\end{equation}
and mNGs exist with the mass linearly depending on $\theta$. However, unlike \textit{Protocol} \romannumeral1, it does not allow for a hierarchical symmetry breaking process, and the symmetry $G_1$ is already broken by $Q^{(1)}$ of order $\mathcal{O}(T)$. 

Both \textit{Protocol} \romannumeral1\ and \romannumeral2\ require the direct access of the Hamiltonian $H_2$ that preserves a non-abelian symmetry $G_2$. This can be particularly challenging for state-of-the-art quantum simulator platforms. In practice, the accessible instantaneous Hamiltonian can take the following form, $H=H_n+H_{n-1}+\cdots+H_0$, where $H_0$ breaks all symmetries. Yet, flipping some of the signs in front of each Hamiltonian can be easily achieved, for instance, by applying single-site $\pi/2$ gates on Rydberg atom quantum simulators~\cite{geier2021floquet}. Next, we introduce two Floquet protocols by combining single-site gates and the many-body Hamiltonian $H$, such that Floquet-engineering mNGs become experimentally feasible.

\textit{Protocol} \romannumeral3.
We consider the 4-step drive
\begin{equation}
    U_F=U_{(+1,+1)}U_{(-1,-1)}U_{(+1,-1)}U_{(-1,+1)}
\end{equation}
where $U_{(\tau_1,\tau_2)}=e^{-i(H_2+\tau_1\cdot H_1+\tau_2\cdot H_0)T},\ \tau_i=\pm1$, $H_2$ preserves $G_2$ symmetry, $H_1$ preserves $G_1\subset G_2$, and $H_0$ breaks all symmetries. Its effective Hamiltonian reads $Q^{(0)}\propto H_2$, $Q^{(1)}\propto iT[H_2,H_1]$, which features a hierarchical symmetry structure, and $G_1$ is broken by $Q^{(2)}$. 
However, $Q^{(1)}$ does not correspond to the generator of $G_2$ and, therefore, the prediction of the mNG mass can be difficult. We resolve this issue by considering the following protocol.

\textit{Protocol} \romannumeral4.
The protocol reads
\begin{equation}
\begin{aligned}
    U_F=&U_{(+1,+1,+1)}U_{(+1,-1,-1)}U_{(-1,-1,-1)}U_{(-1,+1,+1)}\\&U_{(-1,-1,+1)}U_{(-1,+1,-1)}U_{(+1,+1,-1)}U_{(+1,-1,+1)}
\end{aligned}
\end{equation}
where $U_{(\tau_1,\tau_2,\tau_3)}=e^{-i(H_2+\tau_1\cdot H_1+\tau_2\cdot H_1'+\tau_3\cdot H_0)T},\ \tau_i=\pm1$, and the Hamiltonians $H_2$, $H_1$, $H_1'$ and $H_0$ are the same as Eq.~\eqref{eq.protocol}. Its effective Hamiltonian reads, $Q^{(0)}\propto H_2$, $Q^{(1)}\propto iT[H_1,H_1']$, and $Q^{(1)}$ is a generator of $G_2$. This protocol combines the advantages of both \textit{protocol} \romannumeral1\ and \textit{protocol} \romannumeral3, allowing for an exact prediction of mass, while still being experimentally feasible.

In summary, we now provide four different protocols and discuss their advantages and disadvantages, in terms of both their experimental feasibility and the properties of the realized mNGs. Importantly, we propose \textit{protocol} \romannumeral3\ and \textit{protocol} 
 \romannumeral4\  that do not require direct access to non-abelian many-body Hamiltonians, demonstrating the feasibility of Floquet engineered mNGs for state-of-the-art quantum simulator platforms. This variety of different protocols also highlights that our proposed Floquet-engineered mNGs is a fundamental and fairly universal emergent phenomenon.
\subsection{Continuous Driving Protocol}
The protocol proposed in the main text involves discrete drives that can be implemented on digital quantum simulators. For analog platforms, like cold atoms, a continuous driving protocol is preferred. Here, we present one possible continuous protocol that can achieve a similar effective Hamiltonian to generate mNGs.
 
Consider a unitary time evolution operator
\begin{equation}
    U_F=\mathcal{T}e^{-i\int_0^TH(t)dt}\equiv e^{-iTQ}
\end{equation}
where $H(t+T)=H(t)$ is a time-periodic Hamiltonian and $Q$ is the effective Hamiltonian. $Q$ can be obtained by the high-frequency expansion and its lowest two orders read
\begin{equation}
\begin{aligned}
Q^{(0)}\propto K_0,\qquad
Q^{(1)}\propto T\sum_{l=1}^\infty\frac1l[K_l,K_{-l}],
\end{aligned}
\end{equation}
where $K_l$ denotes the Fourier harmonics of $H(t)$, i.e., $H(t)=\sum_lK_l e^{il\omega t}$. Suppose the time-dependent Hamiltonian has the form
\begin{equation}
    H(t)=K_0+K_{1}\cos{\omega t}+K_{1}^{\prime}\sin{\omega t}+K_2\cos{2\omega t}
\end{equation}
where $K_0$ preserves a symmetry group $G_2$, $K_{1}$ and $K_1^{\prime}$ are two non-commuting generators of $G_2$, and $K_2$ further breaks all symmetries. Then the effective Hamiltonian reduces to $Q^{(0)}\propto K_0,\ Q^{(1)}\propto iT[K_{1},K_{1}^{\prime}]$ and higher-order terms break all symmetries. Hence, this protocol can effectively realize mNGs similar to those in the main text but with continuous drives.

\section{Further Analysis for Ferromagnetic Systems}
\subsection{Derivation for the Spin-wave Spectrum, Eq.~\eqref{eq.spf}}
Based on Eq.~\eqref{eq.truancatedH} and $\dot{S}_j^\mu = \{S_j^\mu,Q_{\text{eff}}\}$, one can derive the EOM 
\begin{equation}\label{eq.eomfield}
    \begin{aligned}
        \dot{S}_j^x=&\frac{J_2}{5}[(S_{j-1}^z{+}S_{j+1}^z)S_j^y{-}(S_{j-1}^y{+}S_{j+1}^y)S_j^z]{-}\frac{h_xh_yT}{200}S_j^y,\\
        \dot{S}_j^y=&\frac{J_2}{5}[(S_{j-1}^x{+}S_{j+1}^x)S_j^z{-}(S_{j-1}^z{+}S_{j+1}^z)S_j^x]{+}\frac{h_xh_yT}{200}S_j^x,\\
        \dot{S}_j^z=&\frac{J_2}{5}[(S_{j-1}^y{+}S_{j+1}^y)S_j^x{-}(S_{j-1}^x{+}S_{j+1}^x)S_j^y].
    \end{aligned}
\end{equation}
To derive the mNGs spectrum, we employ a standard linearization method and consider the regime, $S_j^z{\approx} 1$ and $S_j^{x/y}{\ll} 1$, assuming that $h_x$ and $h_y$ have the opposite sign. Neglecting non-linear terms in $S_j^{x/y}$,  we obtain
\begin{equation}\label{lineareom}
    \begin{aligned}
        \dot{S}_j^x=&\frac{J_2}{5}(2S_j^y-S_{j-1}^y-S_{j+1}^y)-\frac{h_xh_yT}{200}S_j^y,\\
        \dot{S}_j^y=&\frac{J_2}{5}(S_{j-1}^x+S_{j+1}^x-2S_j^x)+\frac{h_xh_yT}{200}S_j^x, \\
        \dot{S}_j^z=& 0\ .
    \end{aligned}
\end{equation}
This can be solved analytically by performing a Fourier transform, $S_j^\alpha{=}\sum_q S_q^\alpha e^{i(qa)j}$, $\alpha{\in}\{x, y, z\}$, with $q$ denoting the quasi-momentum and $a$ being the lattice constant. Finally, we get the spin-wave spectrum, Eq.~\eqref{eq.spf}
\begin{equation}\label{eq.sepcturm}
    \omega=\pm\left[\frac{2J_2}{5}(1-\cos{qa})+\frac{|h_xh_y|T}{200}\right].
\end{equation}
In all numerical simulations, we take $a=1$.

\subsection{Oscillating Amplitudes of Quasi-conserved Quantities in Classical Spin Systems}\label{SM.Amplitude}
The oscillatory dynamics of the $z$-component magnetization, shown in Fig.~\ref{fig.dynamicsfield}, cannot be explained by the truncated effective Hamiltonian, Eq.~\eqref{eq.truancatedH}. In Fig.~\ref{fig.Afield}, we show that the oscillating amplitude of the total magnetization in different directions can have different dependence on the drive period $T$. Here we provide one perturbative argument to justify this behavior.

Drive-induced terms $Q^{(2)}$ of order $\mathcal{O}(T^2)$ break the $O(2)$ symmetry along the $z$-axis. Therefore, in the linearized regime where spins are mostly polarized in the $z$ direction, $Q^{(2)}$ effectively generates weak fields in the $(x,y)$-plane. One can approximately describe this situation via the Hamiltonian
\begin{equation}
    H=-J_2\sum_{j}\boldsymbol{S}_j\cdot\boldsymbol{S}_{j+1}-h\left(h_x\sum_j S_j^x+h_y\sum_j S_j^y+h_z\sum_j S_j^z\right),
\end{equation}
where $h_x^2+h_y^2+h_z^2=1$, and $h$ quantifies the strength of the field. In the high frequency regime, we expect $h\sim T$, $h_{x/y}\sim T$ and $h_{z}\sim 1$ for small drive periods $T$. 
The ground state of this Hamiltonian is $\Vec{S}_j=(h_x, h_y, h_z)$ for all sites. One can obtain the linearized EOM 
% \hz{do we require $S_z\to 1$?}\hy{not here}
\begin{equation}\label{eomhoti}
    \begin{aligned}
        \dot{S}_j^x=&J_2[h_y(S_{j-1}^z+S_{j+1}^z-2S_j^z)+h_z(-S_{j-1}^y-S_{j+1}^y+2S_j^y)+h(-h_yS_j^z+h_zS_j^y),\\
        \dot{S}_j^y=&J_2[h_z(S_{j-1}^x+S_{j+1}^x-2S_j^x)+h_x(-S_{j-1}^z-S_{j+1}^z+2S_j^z)+h(-h_zS_j^x+h_xS_j^z),\\
        \dot{S}_j^z=&J_2[h_x(S_{j-1}^y+S_{j+1}^y-2S_j^y)+h_y(-S_{j-1}^x-S_{j+1}^x+2S_j^x)+h(-h_xS_j^y+h_yS_j^x).\\
    \end{aligned}
\end{equation}
By Fourier transforming $S_j^\alpha=\sum_q S_q^\alpha e^{i(qa)j}$ we have
\begin{equation}
    \dot{\boldsymbol{S}}_q=[2J_2(1-\cos{qa})+h]\begin{pmatrix}
    0 & h_z  & -h_y\\
    -h_z & 0  & h_x\\
    h_y & -h_x & 0 \\
    \end{pmatrix}\boldsymbol{S}_q
    \equiv A_q\boldsymbol{S}_q.
\end{equation}
The solution of the ODE reads
\begin{equation}
    \boldsymbol{S}_q(t)=e^{A_q t}\boldsymbol{S}_q(0),
\end{equation}
with the expression
\begin{equation}\label{eq.ev}
   e^{A_q t}=\left(
\begin{array}{ccc}
 {\left(h_y^2+h_z^2\right) \cos \left(\omega t\right)+h_x^2} & {2 h_x h_y
   \sin ^2\left(\frac{1}{2} \omega t\right)+h_z
    \sin \left(\omega t\right)} & {2 
   h_x h_z \sin ^2\left(\frac{1}{2} \omega t\right)-h_y
    \sin \left(\omega t\right)} \\
 {2  h_x h_y \sin ^2\left(\frac{1}{2} \omega t\right)-h_z 
   \sin \left(\omega t\right)} & {h_x^2 \cos \left(\omega t\right)+h_z^2 \cos \left(\omega t\right)+h_y^2} & {2 h_y h_z
   \sin ^2\left(\frac{1}{2} \omega t\right)+h_x
    \sin \left(\omega t\right)} \\
 {2  h_x h_z \sin ^2\left(\frac{1}{2} \omega t\right)+h_y 
   \sin \left(\omega t\right)} & {2  h_y h_z \sin ^2\left(\frac{1}{2} \omega t\right)-h_x 
   \sin \left(\omega t\right)} & {h_x^2 \cos \left(\omega t\right)+h_y^2 \cos \left(\omega t\right)+h_z^2} \\
\end{array}
\right),
\end{equation}
suggesting an oscillatory dynamics with frequency $\omega=2J_2(1-\cos{qa})+h$. It is important to notice that, although the field strength in the $(x,y)$-plane is much smaller than the $z$-field, the oscillations in the three directions couple together and share the same frequency.
However, the oscillation amplitudes are different, if the initial states are almost polarized in the $z$-direction, $S_q^{x/y}\ll1$. More precisely, as long as $S_q^{x/y}$ is of order $\mathcal{O}(T)$ or even smaller, the third column of the matrix in Eq.~\eqref{eq.ev} will dominate the dynamics, and the oscillating amplitudes scale as $A_{x/y}\sim T$ and $A_z\sim T^2$.

\subsection{Nonlinear Effect}

Zooming into Fig.~\ref{fig.spectrumfield}, there exists a very faint line in DSF, which goes beyond the prediction of mNGs obtained by the linearized EOM. We attribute it to the nonlinear effects induced by the Heisenberg model, and the drive-induced higher-order corrections. 

\begin{figure}[h] \includegraphics[width=0.9\linewidth]{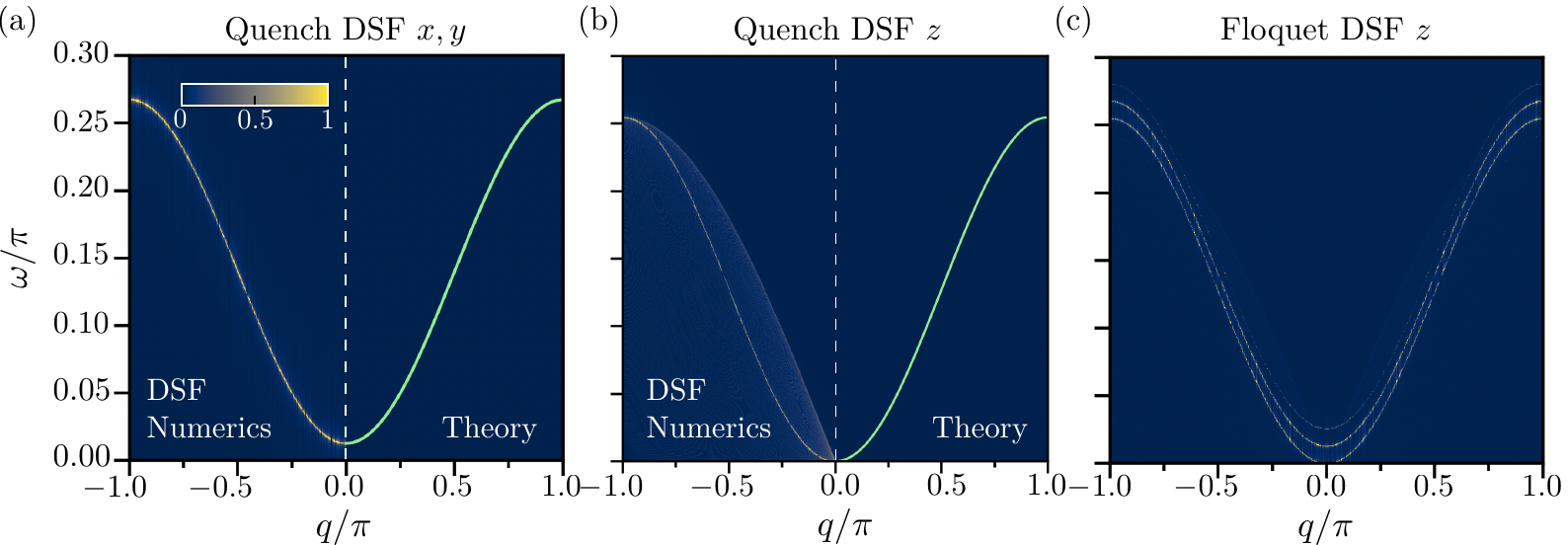}
	\caption{Dynamical structure factor (DSF) over the Brillouin zone. (a) and (b) are obtained from spin-dynamics simulation of a time-independent Hamiltonian, Eq.~\eqref{eq.truancatedH}. (a) contains the contribution of x and y components, and the gapped mNG mode with quadratic dispersion appears, matching well the theoretical prediction. (b) illustrates the DSF of the $z$-component, where a gapless signal is visible due to the nonlinear effect of the Heisenberg model, which also aligns with our analytical calculation. (c) corresponds to the DSF for the $z$-component obtained by the Floquet protocol, Eq.~\eqref{eq.protocol}. Both the gapped mode and the gapless signal can be observed, since magnetizations in three directions are coupled via higher order drive-induced processes.
    }
\label{fig.quenchdsf}
\end{figure}

To show this, we first compare the DSF for the quenched dynamics of $Q_\text{eff}$ and the Floquet protocol.
Fig.~\ref{fig.quenchdsf}(a) depicts the contribution of the $x$ and $y$ components to the DSF, where only the gapped mode is visible. A gapless signature appears in Fig.~\ref{fig.quenchdsf}(b), where the DSF of the $z$ component is plotted. The spin-wave solution of the linearized EOM, Eq.~\eqref{lineareom}, predicts that the $z$-component should remain static. However, when the system is initialized with randomness, spatial fluctuations in z direction indeed become dynamical. Due to total magnetization conservation in the $z$-direction, the excitation spectrum is gapless. 
We now analytically solve for this spectrum by using a perturbative expansion w.r.t.~the linearized solution. As shown on the right side of Fig.~\ref{fig.quenchdsf}(b), it matches well with the most dominant signal obtained by numerical simulations. 

The mean-field solution, i.e., the linearized solution for $q=0$, reads
\begin{equation}
\begin{aligned}\bar{S}^x(t)&=\bar{S}^x(0)\cos{(\omega_{\text{gap}}t)}+\bar{S}^y(0)\sin{(\omega_{\text{gap}}t)},\\
\bar{S}^y(t)&=\bar{S}^y(0)\cos{(\omega_{\text{gap}}t)}-\bar{S}^x(0)\sin{(\omega_{\text{gap}}t)},\\
\bar{S}^z(t)&=\bar{S}^z(0)
\end{aligned}
\end{equation}
where $\omega_{\text{gap}}=|h_xh_y|T/200$. We consider small fluctuations around the mean-field solution as $S_j^\alpha=\bar{S}^\alpha+\delta S_j^\alpha$ with $\delta S_j^\alpha\ll1$. By expanding Eq.~\eqref{eq.eomfield} to the leading order of $\delta S_j^\alpha$ , we obtain 
\begin{equation}
    \begin{aligned}
        \delta\dot{S}_j^x=&\frac{J_2}{5}(2\delta S_j^y-\delta S_{j-1}^y-\delta S_{j+1}^y)-\frac{h_xh_yT}{200}\delta S_j^y-\frac{J_2}{5}\bar{S}_y(2\delta S_j^z-\delta S_{j-1}^z-\delta S_{j+1}^z),\\
        \delta \dot{S}_j^y=&\frac{J_2}{5}(\delta S_{j-1}^x+\delta S_{j+1}^x-2\delta S_j^x)+\frac{h_xh_yT}{200}\delta S_j^x+\frac{J_2}{5}\bar{S}_x(2\delta S_j^z-\delta S_{j-1}^z-\delta S_{j+1}^z), \\
        \delta\dot{S}_j^z=&\frac{J_2}{5}\bar{S}_y(2\delta S_j^x-\delta S_{j-1}^x-\delta S_{j+1}^x)-\frac{J_2}{5}\bar{S}_x(2\delta S_j^y-\delta S_{j-1}^y-\delta S_{j+1}^y) .
    \end{aligned}
\end{equation}

Since the mean-field solution is valid for $\bar{S}^{x/y}\ll1$, the last terms on the right-hand side of the first two rows can be neglected. Then, by performing a Fourier transform, $S_j^\alpha{=}\sum_q S_q^\alpha e^{i(qa)j}$, we obtain $\delta S_q^{x/y}=A_q^{x/y}e^{-i\omega t}$, where $\omega$ is the same as Eq.~\eqref{eq.sepcturm} and $A_q^x=iA_q^y$. Plugging the solution into the EOM of $\delta S_q^z$, we have $\delta \dot{S}_q^z\propto e^{i(\omega-\omega_{\text{gap}})t}$. Therefore, the spectrum of the $z$-components is quadratic and gapless
\begin{equation}
    \omega_z=\omega-\omega_{\text{gap}}=\pm\frac{2J_2}{5}(1-\cos{qa}).
\end{equation}

Furthermore, drive-induced higher-order effects slightly modify the spectrum. Crucially, as is mentioned in the main text, the explicit SB generator of the $O(2)$ symmetry slightly deviates from the $z$-direction due to higher-order corrections, which couple the fluctuations in all three directions. Therefore, in Floquet systems, DSF of the $z$-component (Fig.~\ref{fig.quenchdsf}(c)) exhibits both a gapped and a gapless spectrum. Similarly, the DSF of the $x$ and $y$ components also exhibits a faint gapless spectrum in Fig.~\ref{fig.spectrumfield}.

\subsection{Asymmetric Lineshape and Side Peaks}
The DSF has an asymmetric lineshape, as shown in Fig.~\ref{fig.widthfield}(a). Also, several side peaks can be clearly resolved -- an intriguing and rare non-equilibrium phenomenon in systems without external driving. We conjecture that the side peaks originate from $Q^{(3)}$ of order $\mathcal{O}(T^3)$, a high-order and weak process. Their appearance can be slightly enhanced by the Trotter decomposition being used to approximate the Heisenberg Hamiltonian evolution. To demonstrate this, we compare the numerical simulation generated by $Q_{\text{eff}}$ and different order Trotter decompositions.

\begin{figure}[h] \includegraphics[width=\linewidth]{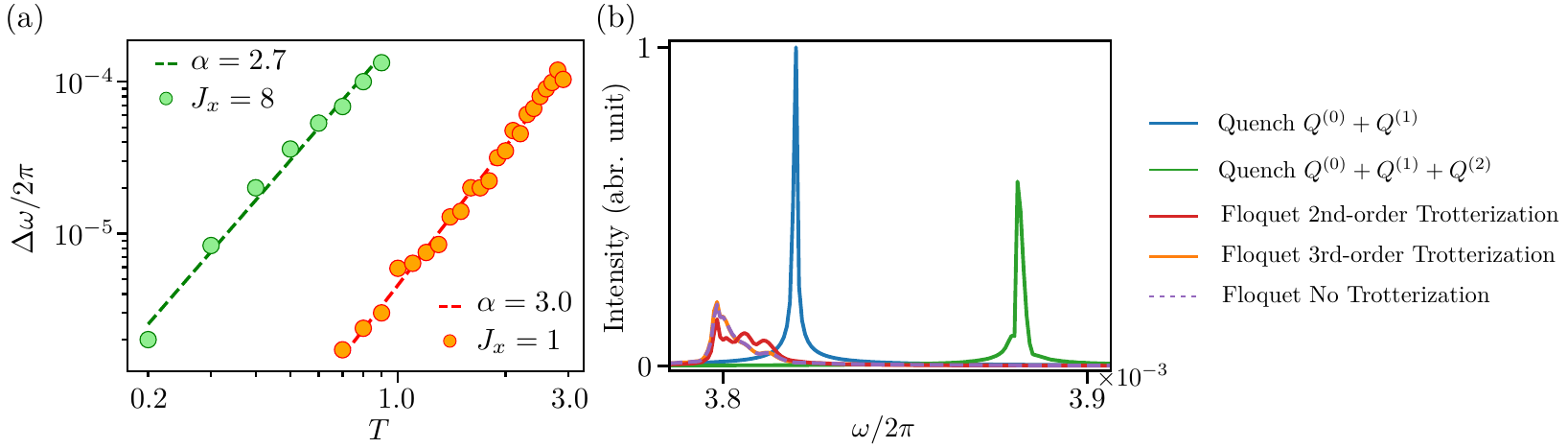}
	\caption{(a) The splitting between the main peak and the highest side peak $\Delta\omega$ exhibits a power-law scaling as $\Delta\omega\sim T^{\alpha}$, $\alpha\approx3$, suggesting that these peaks may be generated by $Q^{(3)}$ of order $\mathcal{O}(T^3)$. (b) Fourier peaks obtained by various numerical methods. We use $T{=}1.2$ and $J_x{=}1$. The lineshape becomes asymmetric once $Q^{(2)}$ is included. Fine structures in the lineshape appear in Floquet systems.
3rd-order Trotterization leads to a Fourier spectrum that is almost identical to those obtained without the Trotterization. 2nd order Trotter slightly enhances the side peaks, although the overall linewidth remains approximately unchanged.  We use $J_2{=}1$, $h_x{=}2$, $h_y{=}{-}2$, $\theta_j{=}0.25\pi$ and $\phi_j$ randomly sampled within $(-\pi/2, \pi/2)$. $100$ ensemble averages are performed for a system size of $N=1000$.
}
\label{fig.lineshape}
\end{figure}

We extract the splitting $\Delta\omega$ between the main peak and the highest side peak and plot it in Fig.~\ref{fig.lineshape}(a) for different driving periods. $\Delta\omega$ exhibits a power-law dependence on the driving period $T$ with the exponent $\alpha\approx3$. 

We then try to reproduce the interesting lineshape by several different numerical methods: 

1. Quenching the initial states with static effective Hamiltonians, truncated at different orders $\mathcal{O}(T)$ and $\mathcal{O}(T^2)$.

2. Floquet driving the system and approximating the Heisenberg Hamiltonian $H_2$ using Trotter decomposition of different orders. Details of this method will be explained later in Sec.~\ref{sec.Trotter}.

3. Floquet driving the system but evolving $H_2$ using a Runge-Kutta solver. 

The results are shown in Fig.~\ref{fig.lineshape}(b). For $Q^{(0)}+Q^{(1)}$, mNGs are exact low-energy quasi-particles and thermal fluctuation broadens the peak in a symmetric way (blue).  The inclusion of $Q^{(2)}$ causes a peak shift; moreover, the lineshape becomes asymmetric (green). 
For Floquet systems without Trotter decomposition (purple), the center of the Fourier peak shifts again and starts to exhibit interesting fine structures, e.g., the shoulder on the right side of the main peak. Although not shown, for a smaller drive period side peaks also arise. Hence, one can conclude that $Q^{(2)}$, which breaks all symmetries, is sufficient to generate an asymmetric lineshape, but the fine structures, e.g., the splitting of the main peak, shoulders or side peaks, originate from $Q^{(3)}$ of order $\mathcal{O}(T^3)$. This also qualitatively explains the scaling exponent in the splitting observed in Fig.~\ref{fig.lineshape}(a).
The 3rd order Trotter approximation leads to almost identical results (orange). Interestingly, the 2nd order Trotter decomposition leads to more visible side peaks. 

\begin{figure}[h] \includegraphics[width=0.4\linewidth]{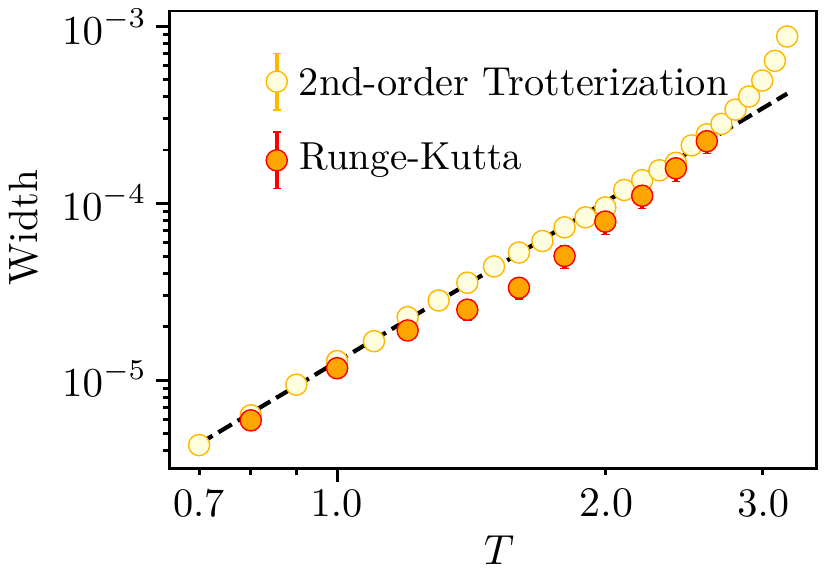}
	\caption{Linewidth of the Fourier spectrum for different simulation methods. The Trotter protocol generates a slightly larger linewidth, but it closely mimics the Runge-Kutta solver. We use $J_2{=}1$, $h_x{=}2$, $h_y{=}{-}2$,  $J_x{=}1$ $\theta_j{=}0.25\pi$ and $\phi_j$ randomly sampled within $(-\pi/2, \pi/2)$. $100$ ensemble averages are performed for a system size of $N=1000$.
}
\label{fig.widthode}
\end{figure}

% The side peaks are more distinguishable as the Trotterization order decreases. 
% \hy{The side peaks can be a higher-order effect, while the thermal fluctuation makes them vague. The Trotterization protocol approximates the Hamiltonian with some integrable steps that process more conserved quantities, resulting in smaller thermal broadening. However, the higher-order Trotterization methods introduce complicated interactions, leading to significant many-body effects and thermal fluctuation. Therefore, we can see distinct side peaks with lower-order Trotterization.}
It is worth pointing out that, although the 2nd Trotterization used in the main text results in a slightly different lineshape, compared to the Runge-Kutta results, the overall linewidths are approximately the same, as shown in Fig.~\ref{fig.widthode}. 

\subsection{Lifetime of Modes with Finite Momenta}

In the main text, we investigate the lifetime of the $q=0$ mode. In Fig.~\ref{fig.widthfield}, we find that the linewidth of the DSF scales as $T^{3}$, and we give one possible explanation using a hydrodynamic theory. Here, we further discuss the lifetime of modes with finite momenta.

For simplicity, we first consider the static system governed by the effective Hamiltonian $Q^{(0)}+Q^{(1)}$. 
In the case of the Heisenberg model with an additional z-field, different $q$ modes have infinite lifetime. Quantum mechanically, this is true since all single-magnon excitations are exact eigenstates of $Q^{(0)}+Q^{(1)}$. Classically, the spin-wave solution of the approximate linearized EOM is in fact an exact solution of non-linear EOM for the classical Heisenberg model. More generally, for the case in the main text of our work and also Ref.~\cite{watanabe2013massive}, where $Q^{(0)}$ preserves a non-Abelian symmetry $G_2$ and $Q^{(1)}$ is proportional to a generator of $G_2$, the infinite lifetime of modes with arbitrary momentum $q$ is also guaranteed. According to the Goldstone Theorem, there exists a gapless excitation for $Q^{(0)}$, $Q^{(0)}|q\rangle=E(q)|q\rangle$, where $E(q)$ denotes the eigenenergy for the single-magnon excitation at a fixed $q$. Due to $[Q^{(0)}, Q^{(1)}]=0$, $Q^{(0)}$ and $Q^{(1)}$ share the same set of eigenstates, which implies that $Q^{(1)}|q\rangle=K(q)|q\rangle$. Therefore, mNGs are also infinitely long-lived with eigenenergy $E(q)+K(q)$. 

This becomes more complicated if the symmetry-breaking term $Q^{(1)}$ is not a generator of $G_2$, hence $[Q^{(0)}, Q^{(1)}]$ no longer vanishes. Therefore, $Q^{(0)}$ and $Q^{(1)}$ do not have simultaneous eigenstates. Construction of the mNG mode, e.g. by using spin-wave approximation, then depends on the specific form of the Hamiltonian. Their lifetime may also depend on $q$ and this certainly goes beyond the scope of this work.

\begin{figure}[h] \includegraphics[width=0.4\linewidth]{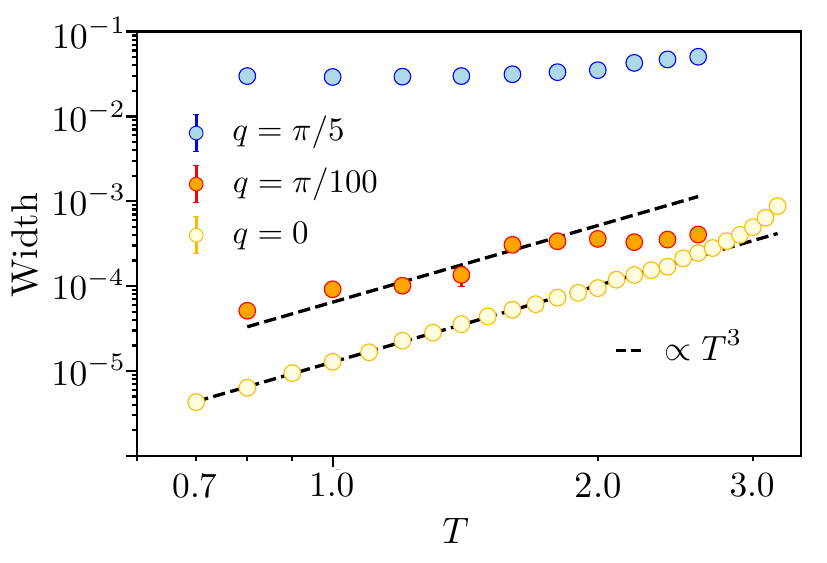}
	\caption{The width for different $q$ modes. The linewidth of the $q=\pi/100$ mode (orange) exhibits a similar $T^3$ scaling behavior in the high-frequency limit as the $q=0$ mode. The black dashed line with scaling $T^3$ is a guide to the eye. The $q=\pi/5$ mode exhibits a broader linewidth, and hence a shorter lifetime. Its lifetime is largely independent of $T$, suggesting that its decay is dominated by $Q^{(0)}$ and the thermal fluctuations in the initial state. We use $J_2{=}1$, $h_x{=}2$, $h_y{=}{-}2$,  $J_x{=}1$ $\theta_j{=}0.25\pi$ for numerical simulations and $\phi_j$ is randomly sampled within $(-\pi/2, \pi/2)$. $100$ ensemble averages are performed for a system size of $N=1000$.
}
\label{fig.widthqmode}
\end{figure}

However, in our work, we always start from an initial state with spatial fluctuations at low but finite temperature. Floquet driving also induces higher-order effects that further destabilize the mNGs. Therefore, we numerically examine the dependence of the linewidth on momentum $q$ of the mode in Fig.~\ref{fig.widthqmode}. We use the same parameters and initial states as the yellow dots in Fig.~\ref{fig.widthfield}(b) and extract the linewidth of two extra modes: one has a momentum close to zero while the other has a large momentum.
The linewidth indeed depends on $q$ in a non-trivial way. For a fixed driving period
$T$, larger $q$ mode exhibits a large linewidth, hence shorter lifetime. The $q=\pi/100$ mode shows a $T^3$ scaling, which is consistent with the hydrodynamics prediction that works particularly well in the long-wavelength limit. For $q=\pi/5$ mode, the linewidth is largely independent of $T$, indicating that mNGs are destabilized by thermal fluctuations in the initial state and $Q^{(0)}$ that is independent of $T$, rather than the driving-induced perturbations. Intuitively, one can understand this behavior by noting that the Floquet protocol only involves local fields or interactions, hence it takes longer time to destroy the long range order, and excitations at short lengthscales are less stable. A deeper understanding of the heating behavior for different $q$ modes would require further investigation, which we leave for future work.

\subsection{Different Scaling Exponent of Linewidth}

In the main text, Fig.~\ref{fig.widthfield}, we find that the linewidth of the DSF scales as $T^{3}$, and we give one possible explanation using a hydrodynamic theory. One key assumption we made is that the attenuation $D_0$, induced by explicit SB, is dominated by $Q^{(2)}$. Together with the scaling of the gap size, one estimates the damping rate as $\Gamma \sim T^3$. 
% This estimation works particularly well when the Hamiltonian $H_0$, that breaks all symmetries, is weak, i.e., $J_x$ is small. 
However, this scaling may not be universal, and it can depend on the driving parameters and the randomness in the initial state ensemble. 

\begin{figure}[h] \includegraphics[width=0.4\linewidth]{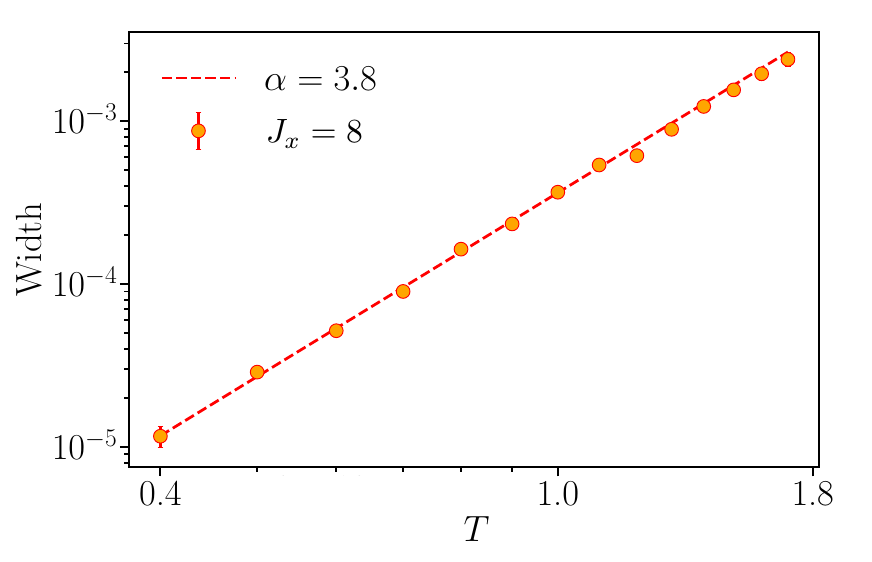}
	\caption{The width $\Gamma$ exhibits a power-law scaling in the high frequency regime; the dashed line with has fitted scaling $\alpha\approx 4$. We use $N{=}1000$, $J_2{=}1$, $h_x{=}2$, $h_y{=}-2$ for numerical simulations. Ensemble averaging over $100$ realizations is performed with $\theta_j{=}0.1\pi$ and azmuthal angles $\phi_j$ randomly sampled within $(-\pi/3, \pi/3)$.
}
\label{fig.widthfieldSM}
\end{figure}

We perform further numerical simulations, using the same parameters as the orange dots in Fig.~\ref{fig.widthfield}(b) but with an initial ensemble with a larger $z-$magnetization, corresponding to a lower temperature. Yet, as shown in  Fig.~\ref{fig.widthfieldSM}, a scaling exponent $\alpha\approx 4$ appears. The reason could be that, although in the high frequency limit the dominant terms in $D_0$ scale as $T^2$, its resulting physical effects may have a more complicated dependence on $T$. Identifying all possible mechanisms for the damping process, and understanding the relation between the hydrodynamics theory and Fermi's Golden rule in this setting, would require substantially more work, which goes beyond the scope of the current study.

\subsection{Thermalization and Hierarchical Symmetry Breaking in Prethermal Plateaus}
\begin{figure}[h] \includegraphics[width=0.4\linewidth]{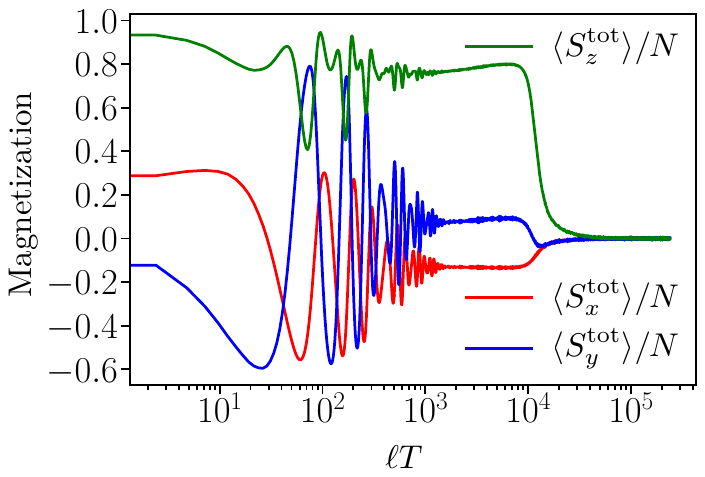}
	\caption{Dynaimics of the 1D driven Heisenberg-type model, Eq.~\eqref{eq.fieldH}, that shows the eventual thermalization. We use $N{=}1000$, $T{=}2.4$ $J_2{=}1$, $h_x{=}2$, $h_y{=}-2$, $J_x{=}5$ for numerical simulations. Ensemble averaging over $100$ realizations is performed with $\theta_j{=}0.1\pi$ and azmuthal angles $\phi_j$ randomly sampled within $(-\pi/3, \pi/3)$.
}
\label{fig.heating}
\end{figure}

Floquet systems generally heat up to infinite temperature, due to the absence of energy conservation. However, this can take a long time, with a timescale that normally increases exponentially with the driving frequency for systems with local interactions~\cite{mori2016rigorous,abanin2015exponentially}. 
In Fig.~\ref{fig.dynamicsfield} in the main text, to show the persistent periodic oscillation of the magnetization we use a small $T$, which indeed significantly delays the onset of heating. During the time window that we can numerically simulate, the system does not heat up to infinite temperature, but stays in a prethermal plateau. Here, instead, we use a smaller driving frequency and stronger symmetry breaking strength $J_x$ to speed up heating. As shown in Fig.~\ref{fig.heating}, after the initial transient oscillation, the system relaxes towards a prethermal plateau. Eventually, magnetization in three directions decays to zero, confirming the eventual heat death. 

{Furthermore, Ref.~\cite{fu2024engineeringhierarchicalsymmetries} predicts a hierarchical symmetry-breaking effect in the prethermal plateau of quasi-conservation laws. To observe this phenomenon, initial states need to be sufficiently random, or high-temperature, but still keeping finite magnetization in the three directions. In the current setup, this effect is difficult to observe since the initial state is close to the ground state and the $Q^{(1)}$ term that breaks the $O(3)$ symmetry only generates a $z$-field, which can be gauged away by performing a rotating frame transformation.  Therefore, the many-body system can not relax to a thermal ensemble corresponding to the truncated Hamiltonian $Q^{(0)}+Q^{(1)}$.
To verify this phenomenon in classical spin systems, we use the following Floquet protocol }

\begin{equation}\label{eq.hsbH}
    U_F=\left(e^{-iH_0\frac{T}{6}}e^{-iH_1\frac{T}{6}}\right)e^{-iH_2\frac{T}{6}}\left(e^{iH_0\frac{T}{6}}e^{iH_1\frac{T}{6}}\right)e^{-iH_2\frac{T}{6}},
\end{equation}

with the Hamiltonians
\begin{equation}
        H_2=-J_2\sum_{j}\boldsymbol{S}_j\cdot\boldsymbol{S}_{j+1},\quad \ H_1=-H_2+J_1\sum_{j\in even}S_j^zS_{j+2}^z,\quad \ 
        H_0=J_x\sum_{j\in odd}S_j^xS_{j+2}^x.
\end{equation}
\begin{figure}[h] \includegraphics[width=0.8\linewidth]{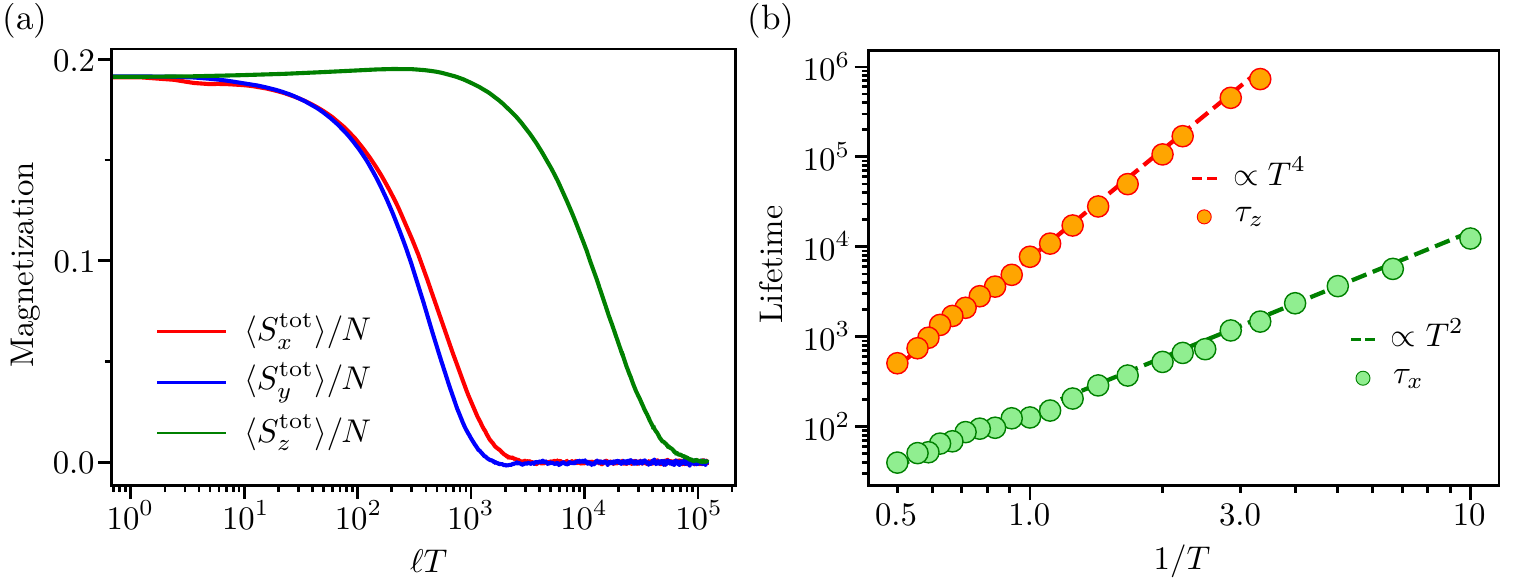}
	\caption{(a) Dynamics of the Floquet system, Eq.~\eqref{eq.hsbH}. The prethermal plateau for magnetization in the $z$ direction survives longer than the other two, showing hierarchical symmetry reduction. (b) We define the lifetime $\tau$ as the time when the magnetizations drop below the threshold value $0.16$. $\tau$ exhibits a power law dependence on the driving period, following the form $\tau\sim T^{-\alpha}$ where $\alpha\approx 2$ for the $x$ (green dots) and $y$ (not shown but nearly coinciding with $x$) components, and $\alpha\approx 4$ for the $z$ (orange dots) components. This scaling is consistent with the Fermi Golden Rule expectation. We use $N{=}100$, $J_2{=}1$, $J_1{=}1$, $J_x{=}1$ in the numerical simulations. Ensemble averaging over $800$ realizations is performed with the initial state where $40\%$ of the sites are prepared as $S_j^x=S_j^y=S_j^z=1/\sqrt{3}$, while the remaining $60\%$ are randomly sampled.
}
\label{fig.hsb}
\end{figure}

{Fig.~\ref{fig.hsb} (a) shows the numerical time evolution of the total magnetization. From a sufficiently random initial ensemble, all components of the magnetization quickly settle into prethermal plateaus 
before eventually heating up to infinite temperature. A hierarchy in the prethermal lifetime is observed, with the $z$-component exhibiting a more stable plateau that persists longer than those of the $x$ and $y$ components. 
Moreover, Ref.~\cite{fu2024engineeringhierarchicalsymmetries} predicts that the prethermal lifetime $\tau$ can be parametrically prolonged by using smaller driving period $T$, $\tau_{x/y}\sim T^{-2}$ and $\tau_{z}\sim T^{-4}$, following a Fermi Golden Rule argument. There, this scaling has been numerically verified in quantum systems of a small size. In Fig.~\ref{fig.hsb} (b) we also observe this scaling law in our classical systems where finite-size effects are negligible. }

\subsection{Classical System with Three-body Interactions}\label{SM.3b}
To realize emergent mNGs, explicit SB processes in $Q^{(1)}$ are not limited to being a generator of the non-Abelian group $G_2$, as long as $Q^{(1)}$ still preserves a subsymmetry of $G_2$.
Here, we present an explicit example that realizes the symmetry-breaking process, $O(3)\to O(2)\to E$, and $Q^{(1)}$ involves three-body interactions which is not a simple generator of the $O(3)$ group. We still observe emergent mNGs with controllable dispersion relation that can be tuned by $T$. However, those mNGs have a shorter lifetime that scales linearly with the driving frequency.

Consider a classical spin chain described by the protocol Eq.~\eqref{eq.protocol} with the Hamiltonians
\begin{equation}\label{eq.3bH}
    \begin{aligned}
         H_2&=-J_2\sum_{j}\boldsymbol{S}_j\cdot\boldsymbol{S}_{j+1},  H_1=J_1\sum_{j}S_{j-1}^yS_{j}^yS_{j+1}^x+S_{j-1}^yS_{j}^xS_{j+1}^y-S_{j-1}^xS_{j}^zS_{j+1}^z,\\
         H_1^{\prime}&=h_y\sum_{j}S_j^y-H_1,
        H_0=J_x\sum_{j}S_j^xS_{j+1}^x.
    \end{aligned}
\end{equation}
We obtain the effective Hamiltonian up to order $\mathcal{O}(T)$ as
\begin{equation}\label{eq.3bheff}
    \begin{aligned}
        Q_{\text{eff}}=-\frac{J_2}{5}\sum_{j}\boldsymbol{S}_j\cdot\boldsymbol{S}_{j+1}+\frac{J_1h_yT}{200}\sum_j(S_{j-1}^xS_{j}^x+S_{j-1}^yS_{j}^y)S_{j+1}^z-S_{j-1}^zS_{j}^zS_{j+1}^z+(S_{j-1}^xS_{j+1}^x+S_{j-1}^yS_{j+1}^y)S_{j}^z.
\end{aligned}
\end{equation}
Following the definition of $Q^{(0)}$ and $Q^{(1)}$ in Eq.~\eqref{eq.effectiveHall}, $Q^{(0)}$ preserves $O(3)$ and $Q^{(1)}$ contains only the exchange terms and the $z$-component terms, therefore reducing the symmetry to $O(2)$ around the $z$-axis. We take $J_2>0$ such that the ground state features the FM order along the positive (negative) $z$-direction if $J_1$ and $h_y$ have the same (opposite) sign. 
The EOM generated by the truncated Hamiltonian Eq.~\eqref{eq.3bheff} read as
\begin{equation}\label{eq.eom3b}
    \begin{aligned}
        \dot{S}_j^x=&\frac{J_2}{5}[(S_{j-1}^z{+}S_{j+1}^z)S_j^y{-}(S_{j-1}^y{+}S_{j+1}^y)S_j^z]{+}\frac{J_1h_yT}{200}[S_{j+1}^yS_{j}^zS_{j+2}^z{+}S_{j-1}^yS_{j}^zS_{j+1}^z{+}S_{j+2}^yS_{j}^zS_{j+1}^z{+}S_{j-2}^yS_{j}^zS_{j-1}^z\\& {+}S_{j}^y(S_{j+1}^zS_{j+2}^z+S_{j-1}^zS_{j+1}^z+S_{j-1}^zS_{j-2}^z){-}(S_{j-1}^xS_{j-2}^x+S_{j-1}^yS_{j-2}^y+S_{j-1}^xS_{j+1}^x+S_{j-1}^yS_{j+1}^y)S_j^y],\\
        \dot{S}_j^y=&\frac{J_2}{5}[(S_{j-1}^x{+}S_{j+1}^x)S_j^z{-}(S_{j-1}^z{+}S_{j+1}^z)S_j^x]{-}\frac{J_1h_yT}{200}[S_{j+1}^xS_{j}^zS_{j+2}^z{+}S_{j-1}^xS_{j}^zS_{j+1}^z{+}S_{j+2}^xS_{j}^zS_{j+1}^z{+}S_{j-2}^xS_{j}^zS_{j-1}^z\\& {+}S_{j}^x(S_{j+1}^zS_{j+2}^z+S_{j-1}^zS_{j+1}^z+S_{j-1}^zS_{j-2}^z){-}(S_{j-1}^xS_{j-2}^x+S_{j-1}^yS_{j-2}^y+S_{j-1}^xS_{j+1}^x+S_{j-1}^yS_{j+1}^y)S_j^x],\\
        \dot{S}_j^z=&\frac{J_2}{5}[(S_{j-1}^y{+}S_{j+1}^y)S_j^x{-}(S_{j-1}^x{+}S_{j+1}^x)S_j^y]{+}\frac{J_1h_yT}{200}[S_{j+2}^z(S_{j+1}^xS_j^{y}{-}S_{j+1}^yS_j^{x}){+}S_{j+1}^z(S_{j-1}^xS_j^y{-}S_{j-1}^yS_j^x)\\&{+}S_{j+1}^z(S_{j+2}^xS_j^y{-}S_{j+2}^yS_j^x){+}S_{j-1}^z(S_{j-2}^xS_j^y{-}S_{j-2}^yS_j^x)].
    \end{aligned}
\end{equation}
We then proceed to show the existence of emergent mNGs by employing a standard linearization method. Similarly, we consider the regime where $S_j^z{\approx} 1$ and $S_j^{x/y}{\ll} 1$ (assuming $J_1$ and $h_y$ have the opposite sign), and neglect non-linear terms in $S_j^{x/y}$, obtaining the linearized EOM
\begin{equation}
    \begin{aligned}
        \dot{S}_j^x=&\frac{J_2}{5}(2S_j^y-S_{j-1}^y-S_{j+1}^y)+\frac{J_1h_yT}{200}(3S_j^y+S_{j-1}^y+S_{j+1}^y+S_{j-2}^y+S_{j+2}^y),\\
        \dot{S}_j^y=&\frac{J_2}{5}(S_{j-1}^x+S_{j+1}^x-2S_j^x)-\frac{J_1h_yT}{200}(3S_j^x+S_{j-1}^x+S_{j+1}^x+S_{j-2}^y+S_{j+2}^x),\\
        \dot{S}_j^z=&0.
    \end{aligned}
\end{equation}
Note that linearization effectively treats the three-body interactions as a field term that points in the $z$-direction.
The spin-wave spectrum now reads 
\begin{equation}\label{eq.sp3b}
    \omega=\pm\left[\frac{2J_2}{5}(1-\cos{qa})+\frac{J_1h_yT}{200}(3+2\cos{qa}+2\cos{2qa})\right].
\end{equation}

A quadratic dispersion appears for modes of long wavelength when $q\rightarrow 0$ and a gap is now opened at $q=0$ of size $7T|J_1h_y|/200$ that is linear in $T$. 

We now confirm the mNG modes and discuss their interplay with hierarchical explicit SB processes, by investigating the spin dynamics generated by the exact Floquet protocol. For the evolution generated by $H_1$ and $H_1^{\prime}$ we numerically integrate the EOMs with the standard Runge-Kutta method. 
Aside from that, we use the same technique as the main text to numerically simulate the system.
\begin{figure}[h] \includegraphics[width=0.5\linewidth]{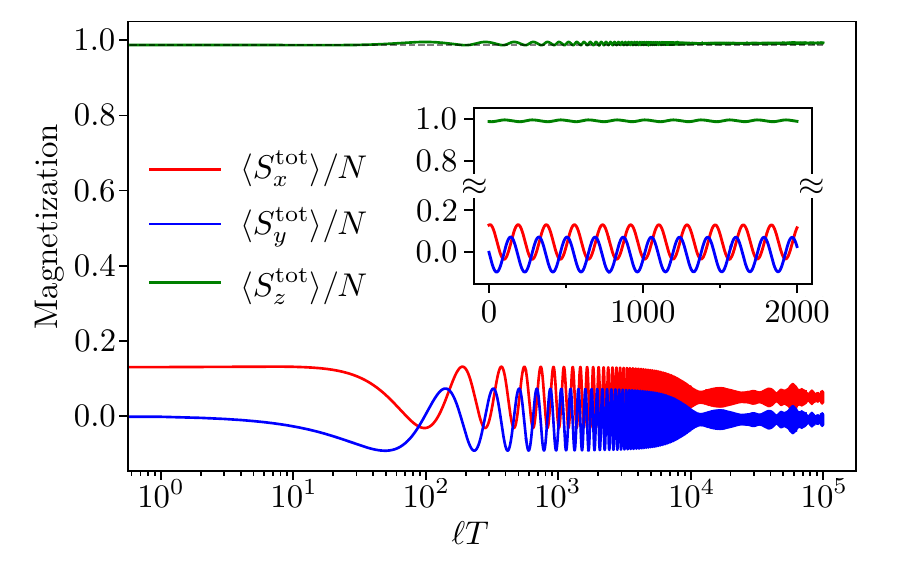}
		\caption{Dynamics of the total magnetization in a 1D driven three-body interaction model, cf.~Eq.~\eqref{eq.3bH}. Magnetization in the $(x,y)$-plane (red and blue) rotates periodically, confirming the existence of mNGs. $\langle S_z^{\text{tot}}\rangle/N$ (green) oscillates on top of the conserved value (dashed line) due to higher order effects. We use a system of size $N=1000$ and parameters $T{=}1$, $J_2{=}1$, $h_y{=}1$, $J_1{=}1$, $J_x{=}1$. We perform ensemble averages over $100$ realizations where initial spins deviate from the $z$-axis by one small angle $\theta_j=0.05\pi$, and their azimuthal angles $\phi_j$ are randomly sampled in the range $(-\pi/3, \pi/3)$. 
}
\label{fig.dynamics3b}
\end{figure}

The initial states are chosen such that $\theta_j=0.025\pi$, and azimuthal angles $\phi_j$ randomly sampled within $(-\pi/3, \pi/3)$. Here $\theta_j$ is close to 0 and hence the condition $S_j^z\approx 1$ is well satisfied. The dynamics of the total magnetizations is shown in Fig.~\ref{fig.dynamics3b}. 
The rotating magnetization in the $(x,y)$-plane (red and blue) suggests the appearance of the mNGs. 

\begin{figure}[h] \includegraphics[width=0.5\linewidth]{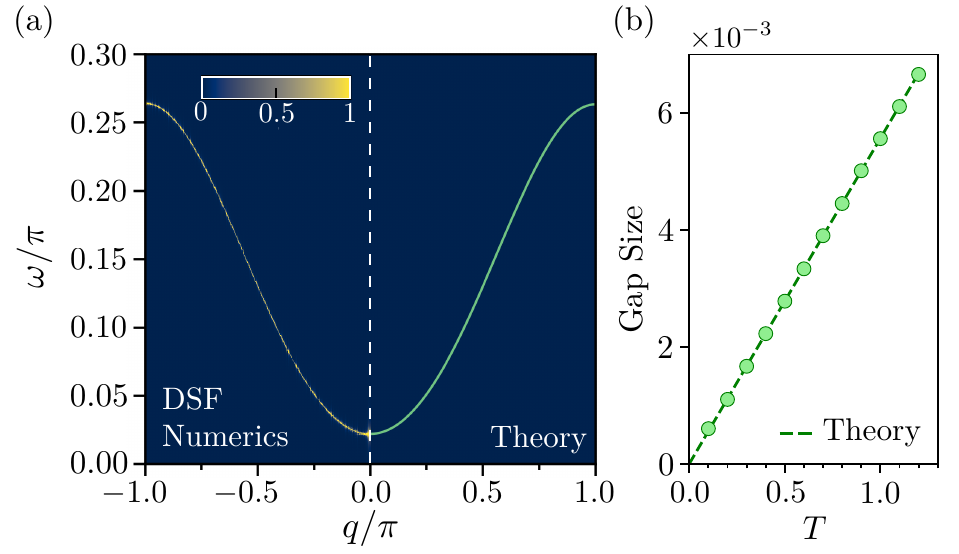}
		\caption{(a) Dynamical structure factor through the first Brillouin zone for the 1D driven three-body interaction model, cf.~Eq.~\eqref{eq.3bH}. Fourier transformation is performed using $\ell=10^4$ Floquet cycles. The left half displays the numerical results, while the right half shows the theoretical predictions. The dispersion spectrum follows the analytical prediction with a gap opened at $q{=}0$. The drive period is taken as $T{=}1$. (b) Linear dependence of $\omega$ versus $T$ for the gap size.
        The system size $N{=}1000$ and the parameters $J_2{=}1$, $h_y{=}1$, $J_1{=}1$, $J_x{=}1$ are used for spin-dynamics simulation.}
\label{fig.spectrum3b}
\end{figure}
We further verify the entire mNG spectrum by analyzing the DSF $\mathcal{S}(q,\omega)$. Fig.~\ref{fig.spectrum3b}(a) depicts  $\mathcal{S}(q,\omega)$ for a fixed $T$, and the spectrum precisely follows our theoretical prediction, Eq.~\eqref{eq.sp3b}. A mNG gap opens at $q=0$ and its size can also be tuned by the drive period, as shown in Fig.~\ref{fig.spectrum3b}(b). 

\begin{figure}[h] \includegraphics[width=0.4\linewidth]{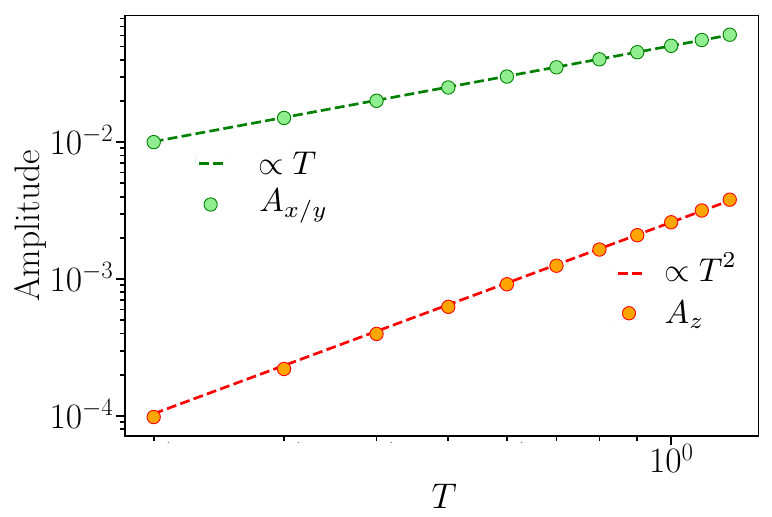}
		\caption{1D driven three-body interaction model, cf.~Eq.~\eqref{eq.3bH}: Scaling of oscillation amplitudes versus $T$. $A_{x/y}\propto T$ and $A_z\propto T^2$ are observed, suggesting that explicit symmetry breaking occurs hierarchically. The initial state is taken to be a spin wave state in the $(x,y)$-plane with $q{=}0.02\pi$, and $\theta_j=0.02\pi$ for all spins. We use $N{=}1000$, $J_2{=}1$, $h_x{=}1$, $J_1{=}1$ and $J_x{=}1$.
}
\label{fig.A3b}
\end{figure}

Similar behavior in the oscillating amplitudes ($A_{x/y}\sim T$, $A_z\sim T^2$) also appears here, see Fig.~\ref{fig.A3b}.

\begin{figure}[!h] \includegraphics[width=0.8\linewidth]{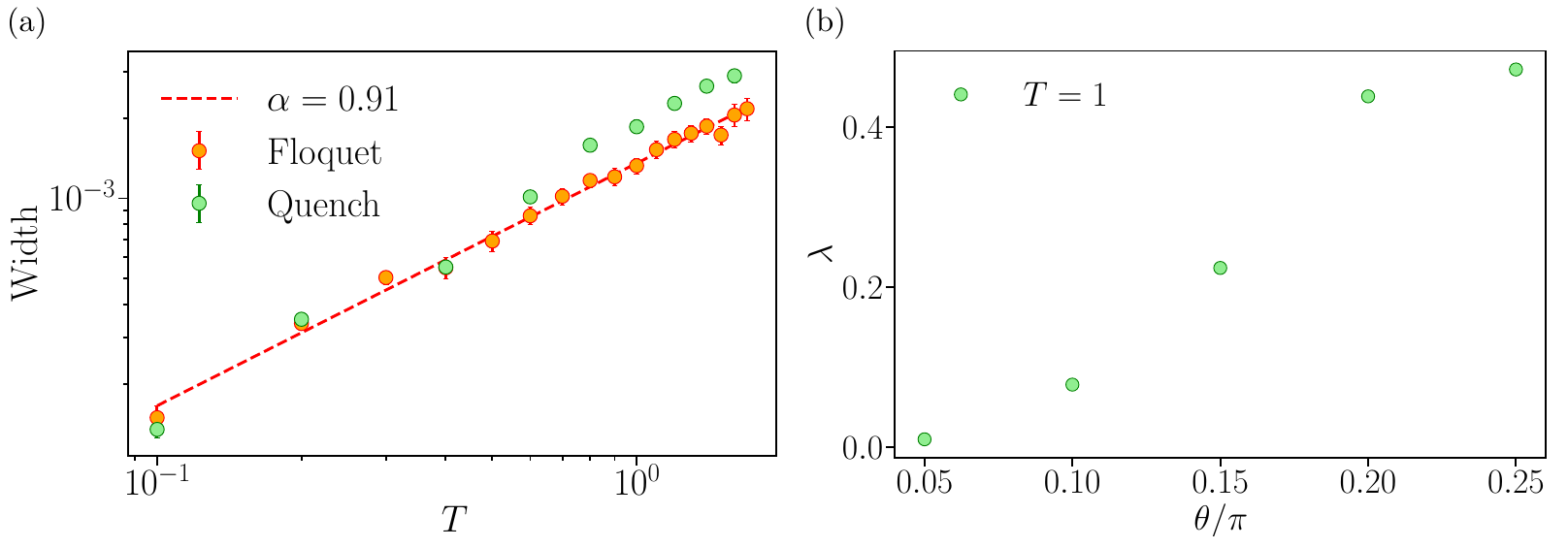}
% The above line can define the position of the figure, and change its width.
		\caption{1D driven three-body interaction model, cf.~Eq.~\eqref{eq.3bH}: (a) The linewidth $\Gamma$ exhibits a power-law scaling in the high-frequency regime with the exponent close to $1$. The results of the quenched protocol closely mimic the Floquet protocol in the high-frequency regime. Floquet protocol generates a smaller linewidth for large $T$. We used $\theta_j{=}0.1\pi$. (b) The monotonic dependence of the ratio of linewidth to gap size, $\lambda$, on initial state parameter $\theta$. The ratio $\lambda$ can be suppressed by applying smaller spatial randomness. Various $\theta_j=\theta$ are used for characterizing the spatial randomness. We use $N{=}1000$, $T{=}1$, $J_2{=}1$, $h_x{=}1$, $J_1{=}1$ and $J_x{=}1$. $100$ ensemble averages are performed with azimuthal angles $\phi_j$ randomly sampled within $(-\pi/3, \pi/3)$.
  % \hz{main message missing!} 
}
\label{fig.width3b}
\end{figure}

For even longer times, the damping of mNG modes broadens the spectrum and induces a finite linewidth $\Gamma$ in the DSF. We extract $\Gamma$ for $q=0$ by performing Fourier transform on the dynamics of the total magnetization components over time, and plot it in Fig.~\ref{fig.width3b} for varying $T$. The linewidth is determined by averaging the full-width at $10\%$, $15\%$, and $20\%$ of the maximum intensity of the Fourier peak, with their standard deviation serving as an error bar. $\Gamma$ exhibits a power-law scaling $T^{\alpha}$. However, in contrast to Fig.~\ref{fig.widthfield}, here the scaling exponent is about 1, cf.~Fig.~\ref{fig.width3b}, and hence mNGs are not very robust. Such a scaling implies that a $Q^{(1)}$ of order $\mathcal{O}(T)$ can destabilize mNGs, even if it preserves the $O(2)$ symmetry. The approximation that one can treat the three-body interaction as an effective $O(3)$ generator fails after a time scale that scales as $\mathcal{O}(T^{-1})$. Therefore, the damping of mNGs becomes notable shortly after the failure of linearization. We simulate the spin dynamics by using the truncated Hamiltonian, Eq.~\eqref{eq.3bheff}, and the results (green dots) are depicted in Fig.~\ref{fig.width3b}(a). In the high frequency regime, the linewidth for this quenched protocol closely mimics the Floquet results, confirming that $Q^{(1)}$ indeed dominates the damping of mNGs. For slightly larger drive period, the Floquet protocol generates a smaller linewidth, i.e., a longer lifetime. It essentially implies that $Q^{(2)}$ indeed stabilizes these mNGs, an counter-intuitive observation since higher order terms normally speed up heating and hence shorten the lifetime of quasi-particles. 

The linear dependence of the linewidth $\Gamma$ and gap size $\Delta$ indicates that the ratio $\lambda\equiv\Gamma/\Delta$ is largely independent of $T$, and hence one cannot stabilize mNGs by increasing the driving frequency. However, one can still control this ratio by employing an initial ensemble with less spatial randomness, or at lower temperatures.
To show this, we prepare the initial state ensemble as in the main text, and change $\theta$ to vary the spatial randomness. As shown in Fig.~\ref{fig.width3b}(b), the ratio $\lambda$ can be systematically reduced upon using smaller spatial randomness. Therefore, the mNGs are well-defined quasi-particles given sufficiently low initial state temperature.

\subsection{Ferromagnetic Quantum Spin Model}
 In the main text, we use a classical spin model to verify the existence of mNGs. However, the Floquet protocol equally applies to both quantum and classical systems. Therefore, we expect similar dynamical phenomena in the quantum system for the FM case. We verify the existence of magnon excitations by applying the protocol Eq.~\eqref{eq.protocol} to a quantum many-body system of spin-$\frac12$, with the Hamiltonian 
\begin{equation}\label{eq.quantumH}
    \begin{aligned}
         H_2=-J_2\sum_{j=1}^{N}{\sigma}_j^x{\sigma}_{j+1}^x+{\sigma}_j^y{\sigma}_{j+1}^y+{\sigma}_j^z{\sigma}_{j+1}^z, \ \ H_1=-h_x\sum_{j=1}^{N}{\sigma}_j^x, \ \
         H_1^{\prime}=-h_y\sum_{j=1}^{N}{\sigma}_j^y, \ \ H_0=J_x\sum_{j=1}^{N}{\sigma}_j^x{\sigma}_{j+1}^x,
    \end{aligned}
\end{equation}
where we use Pauli operators. 
We use Quspin~\cite{marinquspin} for numerical simulation and the system initially has all spins pointing up. Then we perform a single-site rotation on all spins around the $x$-axis by $e^{-i\sum_j\sigma_j^x \theta_j/2}$, where the angle $\theta_j$ is chosen in a small range, yielding a sufficiently large magnetization density in $z$-direction and non-vanishing initial magnetization in the $(x,y)$-plane. The spatial randomness can mimic the thermal fluctuation that may exist in real quantum simulator platforms.
\begin{figure}[h] \includegraphics[width=\linewidth]{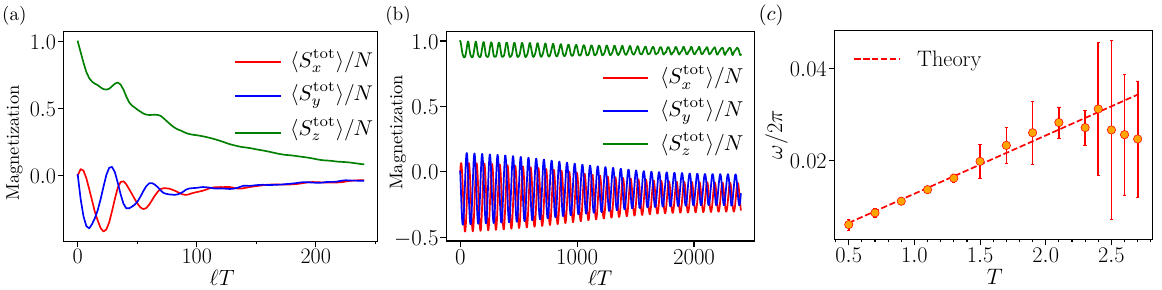}
% The above line can define the position of the figure, and change its width.
	\caption{(a) Dynamics of the total magnetization of the quantum spin chain, Eq.~\eqref{eq.quantumH}, with $T=2.4$. Magnetization in the $(x,y)$-plane (red and blue) rotates at early times, and eventually decays. (b) Dynamics of the total magnetization with a smaller driving period $T=1.2$. Persistent periodic rotation of magnetization in the $(x,y)$-plane (red and blue) confirms the existence of mNGs. (c) Linear dependence of the gap size versus $T$. We use a system of size $N=20$ and parameters $J_2{=}1$, $h_x{=}2$, $h_y{=}-2$, $J_x{=}1$. We perform ensemble averages over $100$ realizations where all spins rotate around the $x$-axis by $e^{-i\sum_j\sigma_j^x \frac{\theta_j}{2}}$ with $\theta_j$ randomly sampled within the range $(-\pi/100,\pi/100)$.
}
\label{fig.quantumfm}
\end{figure}

Fig.~\ref{fig.quantumfm}(a) and (b) illustrate the dynamics of total magnetization. The rotating magnetization around the $z$-axis (red and blue) confirms the presence of the mNG modes, which eventually damp out. The gap of the mode, corresponding to the oscillatory frequency of the magnetization, depends linearly on drive period $T$, as shown in Fig.~\ref{fig.quantumfm}(c). The theoretical calculation may fail for large $T$ that fall outside the high frequency regime. These dynamical behaviors, analogous to those observed in the classical system, offer compelling evidence and support the validity of our analysis in the quantum system.

\section{Massive Nambu-Goldstone Modes in Floquet Anti-ferromagnetic Systems}

We have provided a comprehensive analysis of FM in the main text, categorized as Case 1. For Case 2, we investigate the properties of the mNG modes in the Floquet system with antiferromagnetic (AFM) interaction, $J_2<0$. Significant distinctions between the two cases can be observed. For AFM, if we consider only the first two orders of the effective Hamiltonian, there exists a gapless Goldstone mode with linear dispersion and a gapped mNG with
quadratic dispersion ~\cite{watanabe2013massive}. Taking higher-order terms into account, the gapless mode disappears and the spectrum of the gapped mode is slightly modified. The gap linearly depends on the drive period $T$ in the high-frequency regime. Furthermore, we note that the stability of the gapped mode is quite different from the FM case.
\subsection{Spin-wave Theory}
We first derive the mNG properties theoretically, before showing the numerical verification.
To begin with, we focus on the truncated Hamiltonian, Eq.~\eqref{eq.truancatedH}, labeling $J\equiv J_2/5$ and the  magnetic field strength along the $z$-direction by $h_z\equiv \frac{h_xh_yT}{200}$
\begin{equation}\label{eq.afmquench}
    Q_{\text{eff}}=-J\sum_{j}S_j^xS_{j+1}^x+S_j^yS_{j+1}^y+S_j^zS_{j+1}^z+h_z\sum_j S_j^z,
\end{equation}
where $J<0$ for AFM interaction. If $h_z=0$, there is one NG mode with linear dispersion. For non-zero $h_z$ the ground state of the system features all spins having the same $z$-component with $S_j^z=\frac{h_z}{4J}$ and being antiparallel in $(x,y)$-plane for $|h_z|<-4J$. For $|h_z|>-4J$, the ground state state is just the FM state along the $z$-axis. We focus on the former situation, since in our Floquet system the effective $z$-field strength linearly scales with $T$, which is a perturbatively small quantity. The Hamiltonian leads to the EOM
\begin{equation}\label{eq.eomafm}
    \begin{aligned}
        \dot{S}_j^x=&J[(S_{j-1}^z+S_{j+1}^z)S_j^y-(S_{j-1}^y+S_{j+1}^y)S_j^z]-h_zS_j^y,\\
        \dot{S}_j^y=&J[(S_{j-1}^x+S_{j+1}^x)S_j^z-(S_{j-1}^z+S_{j+1}^z)S_j^x]+h_zS_j^x,\\
        \dot{S}_j^z=&J[(S_{j-1}^y+S_{j+1}^y)S_j^x-(S_{j-1}^x+S_{j+1}^x)S_j^y].
    \end{aligned}
\end{equation}
Without loss of generality, we assume that SSB happens in the $x$-direction. We linearize the EOM around the ground state by plugging $S_j^x=(-1)^jS_{\parallel}+\delta S_j^x$, $S_j^y=\delta S_j^y$ and $S_j^z=S_{\perp}+\delta S_j^z$ into Eq.~\eqref{eq.eomafm} and keep terms up to the linear order of $\delta S_j^{\alpha}$, where $\delta S_j^{\alpha} \ll 1,\; \alpha\in\{x,y,z\}$, $S_{\perp}=\frac{h_z}{4J}$ and $S_{\perp}^2+S_{\parallel}^2=1$. Then we obtain
\begin{equation}
    \begin{aligned}
        \delta\dot{S}_j^x=&\frac{h_z}{4}(2\delta S_j^y-\delta S_{j-1}^y-\delta S_{j+1}^y)-{h_z}\delta S_j^y,\\
        \delta\dot{S}_j^y=&(-1)^{j+1}JS_{\parallel}(\delta S_{j-1}^z+\delta S_{j+1}^z+2\delta S_j^z)+\frac{h_z}{4}(\delta S_{j-1}^x+\delta S_{j+1}^x-2\delta S_j^x)+h_z\delta S_j^x,\\
        \delta\dot{S}_j^z=&(-1)^{j}JS_{\parallel}(\delta S_{j-1}^y+\delta S_{j+1}^y+2\delta S_j^y).
    \end{aligned}
\end{equation}
Note, since the ground state has two-site translation symmetry, we apply the Fourier transform to even and odd sites separately, $S_{j\in \text{even/odd}}^\alpha=\sum_q S_{e/o}^\alpha(q) e^{i(qa)j}$, $\alpha\in\{x, y, z\}$. For each $q$ we can solve the linearized EOM analytically and obtain the dispersion relation 
\begin{equation}\label{eq.disafm}
    \begin{aligned}
       \omega_1 =& \pm\left| \sin \left(\frac{qa}{2}\right)\right|  \sqrt{8 J^2 S_{\parallel}^2 (\cos
   (qa)+1)-\frac{h_z^2}{2} (\cos (qa)-1)},\\
   \omega_2=&\pm\cos \left(\frac{qa}{2}\right) \sqrt{\frac{h_z^2}{2} (\cos (qa)+1)-8 J^2 S_{\parallel}^2 (\cos (qa)-1)},
    \end{aligned}
\end{equation}
where one is a gapless mode with linear dispersion $\omega_1=\pm2JS_{\parallel}|qa|+\mathcal{O}(|qa|^3)$ for small $q$, created by the SSB of $S_z^{\text{tot}}=\sum_jS_j^z$, while the explicit SB of $S_x^{\text{tot}}=\sum_jS_j^x$ and $S_y^{\text{tot}}=\sum_jS_j^y$ creates the gapped one with quadratic dispersion $\omega_2= \pm h_z+\mathcal{O}(|qa|^2)$. The gap size equals the field strength, $h_z$, similar to the gapped mode in FM systems.

In principle, linearization can be carried out through different methods. Here, we present an alternative way to derive the dispersion for AFM which leads to the same dispersion.
We parameterize the spin variables as $S_j^x=\sin{\theta_j}\cos{\phi_j}$, $S_j^x=\sin{\theta_j}\sin{\phi_j}$, $S_j^x=\cos{\theta_j}$, and linearize the EOM of $\theta$ and $\phi$ to leading order
\begin{equation}
    \begin{aligned}        \delta\dot{\theta}_j=&J\sin{\theta_0}(\delta\phi_{j-1}+\delta\phi_{j+1}-2\delta\phi_{j}),\\
        \delta\dot{\phi}_j=&\frac{J_2}{\sin{\theta_0}}[2\delta\theta_{j}-\cos{2\theta_0}(\delta\theta_{j+1}+\delta\theta_{j-1})],
    \end{aligned}
\end{equation}
where $\cos{\theta_0}=h_z/4J$. This EOM can be solved by performing the Fourier transform to the angle variables $\delta\theta_{j\in \text{even/odd}}=\sum_q \delta\theta_{e/o}(q) e^{i(qa)j}$ and $\delta\phi_{j\in \text{even/odd}}=\sum_q \delta\phi_{e/o}(q) e^{i(qa)j}$, leading to two modes
\begin{equation}
    \begin{aligned}
       \omega_1^{\prime} =& \pm\left| \sin \left(\frac{qa}{2}\right)\right|  \sqrt{8 J^2(1-\cos{2\theta_0}\cos{qa})},\\
   \omega_2^{\prime}=&\pm\cos \left(\frac{qa}{2}\right) \sqrt{8 J^2(1+\cos{2\theta_0}\cos{qa})},
    \end{aligned}
\end{equation}
which reproduce the result Eq.~\eqref{eq.disafm}.

\subsection{Numerical Verification of Massive Goldstone Modes}
Here we certify the mNGs via numerical simulation. The AFM ground state of the Hamiltonian Eq.~\eqref{eq.afmquench} can be formulated as
\begin{equation}
\label{eq.AFMgs}
    \cos{\theta_j}=-\frac{h_xh_yT}{160J_2},\  \text{for all}\ j;\ \ \  \phi_{j}=0, \ \text{for}\ j\in \text{even}, \phi_{j}=\pi,\  \text{for}\ j\in \text{odd}.
\end{equation}
% $\cos{\theta_j}=-\frac{h_xh_yT}{160J_2}$ for all $j$, and $\phi_{j}=0$, for $j\in \text{even};\phi_{j}=\pi$, for $j\in \text{odd}$; 
We prepare the initial state as follows: first, the system is initialized as all spins deviate from the $z$-axis by a small angle $\delta\theta_j=0.01\pi$ and $\delta\phi_j$ are randomly sampled within the range $(-\pi/3, \pi/3)$; then all spins are rotated by $\theta_j$ and $\phi_j$ in Eq.~\eqref{eq.AFMgs}. In this way, we generate the initial ensemble with spatial randomness on top of the ground state.
% We will decribe the initial ensemble using $\delta\theta_j$ and $\delta\phi_j$ in the following text.

\begin{figure}[h] \includegraphics[width=\linewidth]{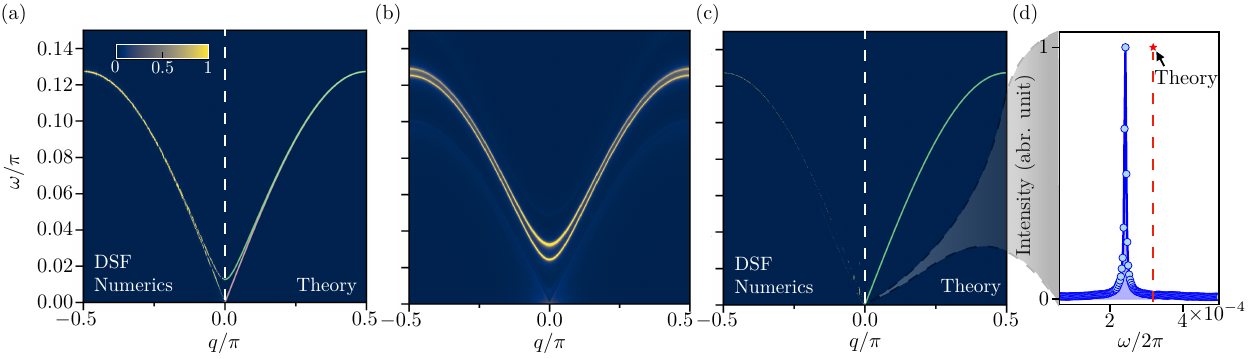}
		\caption{Dynamical structure factor (DSF) through the Brillouin zone of AFM model. (a) Obtained from spin-dynamics simulation of a time-independent Hamiltonian, Eq.~\eqref{eq.afmquench}, with parameters $J{=}{-}1/5$, $h_z{=}1/25$. A gapped mode with quadratic dispersion and a gapless mode with linear dispersion appear in accordance with the theoretical prediction. (b) Floquet spin-dynamics simulation with $T{=}2$, $J_2{=}{-}1$, $h_x{=}2$, $h_y{=}2$ and $J_x{=}1$, which gives the same parameters as (a). However, this spectrum is very different from (a) due to significant higher-order perturbations that are not involved in $Q_{\text{eff}}$. Fourier transformation is performed by using $\ell{=}10^4$ Floquet cycles. (c) Obtained from spin-dynamics simulation of Eq.~\eqref{eq.fieldH} with $T{=}0.4$, $J_2{=}{-}1$, $h_x{=}1$, $h_y{=}1$ and $J_x{=}1$. The gapped mode still matches our theory, while the gapless mode disappears.  Fourier transformation is performed by using $\ell{=}5\times10^4$ Floquet cycles. The Fourier density of $q=0$ is illustrated in (d), where the gap is clearly visible. System size is $N=1000$. Ensemble averaging over $100$ realizations is performed with $\delta\theta_j{=}0.01\pi$ and $\delta\phi_j$ randomly sampled within $(-\pi/3,\pi/3)$ for (a) and (b), while $\delta\phi_j$ randomly sampled within $(-\pi/6,\pi/6)$ for (c).}
\label{fig.spectrumafm}
\end{figure}

To verify Eq.~\eqref{eq.disafm}, we first simulate the quenched dynamics with the time-independent Hamiltonian Eq.~\eqref{eq.afmquench}. Numerically we integrate the EOM Eq.~\eqref{eq.eomafm} via the standard Runge-Kutta method. The DSF computed through the first Brillouin zone is shown in Fig.~\ref{fig.spectrumafm} (a): numerically we observe two modes (dashed lines on the left side of panel (a)) that match well with our theoretical prediction (solid lines on the right side). A mNG gap opens at $q=0$ corresponding to the oscillatory frequency of the total magnetization.

For the Floquet system, drive-induced higher-order terms break the $O(2)$ symmetry, and hence the ground state degeneracy is lifted. Therefore, we have to find the ground state of the effective Hamiltonian $\Tilde{Q}_{\text{eff}}=Q^{(0)}+Q^{(1)}+Q^{(2)}$ truncated at order $\mathcal{O}(T^2)$. $Q^{(2)}$ takes a complicated form
\begin{equation}\label{hoti}
    \begin{aligned}
         Q^{(2)}{=}&\frac{T^2}{4000}([H_0{+}H_1{+}H_1^{\prime},[H_1,H_1^{\prime}]]{-}[2H_1{-}2H_2{+}H_1^{\prime},[H_2,H_1]]{+}2[[H_0,H_1{+}H_1^{\prime}],H_2]{-}[3H_1{-}2H_2{+}2H_1^{\prime},[H_2,H_1^{\prime}]]),\\
         {=}&{-}\frac{T^2}{4000}\sum_jh_xh_y(h_xS_j^y{-}h_yS_j^x){-}J_xh_xh_y S_j^y(S_{j{+}1}^x+S_{j{-}1}^x){+}2J_xh_yJ_2[S_j^y(S_{j{+}1}^x{+}S_{j{-}1}^x)^2\\&{-}(S_{j{+}1}^x{+}S_{j{-}1}^x)S_j^x(S_{j{+}1}^y{+}S_{j{-}1}^y){+}(S_{j{+}1}^z{+}S_{j{-}1}^z)S_j^z(S_{j{+}1}^y{+}S_{j{-}1}^y){-}(S_{j{+}1}^z{+}S_{j{-}1}^z)S_j^y(S_{j{+}1}^z{+}S_{j{-}1}^z)],
\end{aligned}
\end{equation}
and determining analytically its effect on the ground state is quite cumbersome. Instead, we numerically minimize the energy $\Tilde{Q}_{\text{eff}}$, by constraining the optimization parameter space within the state manifold with 2-site translation invariance. Note, the optimized state may not be exact but should be close to the true ground state of $\Tilde{Q}_{\text{eff}}$. Also, the optimized state reduces to the AFM ground state Eq.~\eqref{eq.AFMgs}, analytically obtained for $Q^{(0)}+Q^{(1)}$, if we neglect the effect of $Q^{(2)}$. A small deviation from this state appears as $T$ increases. This optimization is not necessary for detecting the mNGs, as long as $T$ is sufficiently small. However, this optimization can generate a clearer signal of DSF as shown in Fig.~\ref{fig.spectrumafm}. 

\begin{figure}[h] \includegraphics[width=0.7\linewidth]{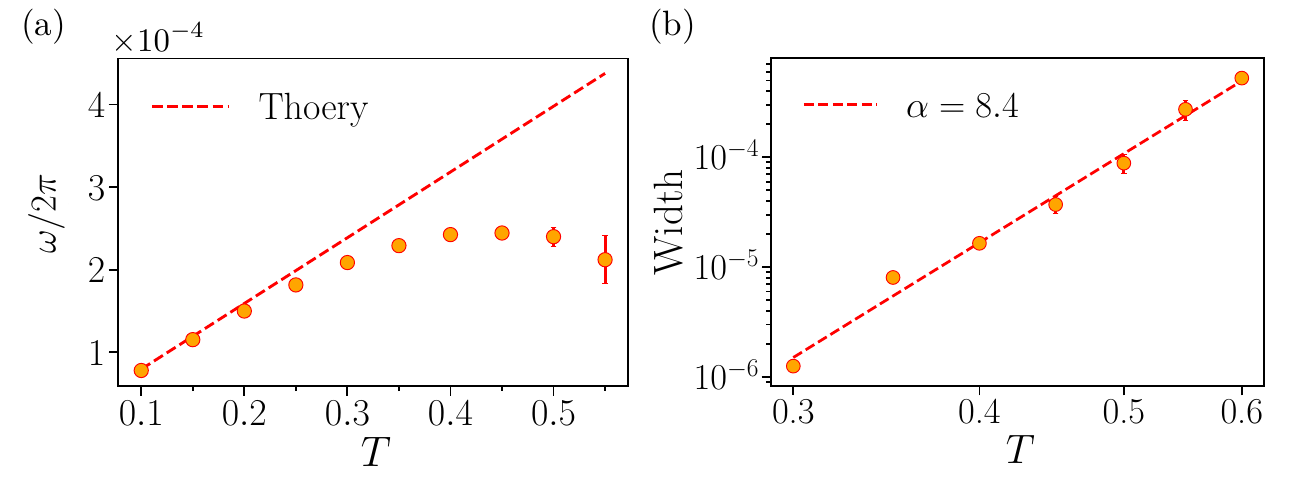}
	\caption{(a) Linear dependence of $\omega$ versus $T$ for the gap size. Deviation between the numerical and theoretical results reduces for higher driving frequencies.  (b) The width $\Gamma$ exhibits a power-law scaling in the high frequency regime with the scaling exponent  $\alpha\approx8$. We use $N=1000$, $J_2=-1$, $h_x=1$, $h_y=1$ and $J_x=1$.  Ensemble averaging over $100$ realizations is performed with $\delta\theta_j{=}0.01\pi$ and $\delta\phi_j$ randomly sampled within the range $(-\pi/6,\pi/6)$.
}
\label{fig.afm1}
\end{figure}

We now simulate the Floquet protocol that generates 
the effective Hamiltonian Eq.~\eqref{eq.afmquench}, with the same parameters as used in Fig.~\ref{fig.spectrumafm}(a). Similarly, we introduce randomness to the optimized state, and obtain the DSF as shown in Fig.~\ref{fig.spectrumafm}(b). The spectrum is drastically different from panel (a). This happens because, in order to match the Hamiltonian parameters we use a large drive period $T$, which causes non-negligible higher-order effects. In Fig.~\ref{fig.spectrumafm}(c) we use a higher driving frequency to suppress these higher order effects, and as shown in panel (d) one mNG mode appears with a finite gap opened at $q=0$. We compare these numerical results with our theoretical prediction (red) in panel (d), where a small deviation appears. As shown in Fig.~\ref{fig.afm1}(a), this discrepancy can be systematically suppressed by reducing the drive period. 

In Fig.~\ref{fig.afm1}(b), we plot the linewidth $\Gamma$ for $q = 0$ by performing Fourier transform to the dynamics of total magnetizations over time, for varying $T$. The linewidth is determined by averaging the full-width at $10\%$, $15\%$, and $20\%$ of the maximum intensity of the Fourier peak, with their standard deviation serving as the error bar. Interestingly, $\Gamma$ exhibits a power-law scaling $T^{\alpha}$ with a large scaling exponent $\alpha\approx 8$, suggesting that the lifetime of the mNG mode can be notably changed by slightly tuning the driving frequency.

We also note that the gapless NG mode no longer exists in the current Floquet setting, implying that its robustness is worse than the gapped mNGs. It remains an interesting and open question how to stabilize gapless modes in Floquet systems.

\section{Explicit Breaking of Abelian Symmetries}
In the main text, we focus on the appearance of mNGs due to the breaking of non-abelian symmetries. In fact, mNGs can also exist in systems with an approximate
abelian continuous symmetry.
However, in this case, mass of mNGs becomes model-dependent and its determination can involve complicated calculations, both numerically and analytically.

For illustration, we consider the 2D antiferromagnetic transverse-field XXZ model on a square lattice, described by the Hamiltonian 
\begin{equation}
    H=\sum_{\langle i,j\rangle}J_\perp (S_i^xS_j^x{+}S_i^yS_j^y){+}J_zS_i^zS_j^z{+}h\sum_j S_j^x,
\end{equation}
with the $U(1)$ symmetry along $z$-direction. When $J_z/J_\perp < 1$ and $h=0$, its ground state features easy-plane antiferromagnetic order, and there is a gapless NG mode with linear dispersion. A finite transverse field degrades the symmetry, $U(1)\to Z_2$, and a mNG mode appears with quadratic dispersion. The gap size can be analytically determined by linear spin-wave theory, exhibiting a complicated dependence on the parameters of the Hamiltonian, see details in Ref.~\cite{kar2017magnon}.
One can further apply an additional field along the z-axis to break the remaining $Z_2$ symmetry, which simply induces an overall energy shift to the mNG spectrum.

In our work, we mainly focus on the non-abelian symmetry and the mass of mNGs is determined by the non-abelian group structure. Therefore, our Floquet protocol is generally applicable to different physical systems of interest and the prediction of the mass is less model-specific.

\section{Trotterized Simulation Protocol}
\label{sec.Trotter}
To achieve an efficient long-time simulation of the classical spin dynamics, we use the Trotterization method for the evolution of the Heisenberg model $H=-J\sum_j \boldsymbol{S}_j\cdot\boldsymbol{S}_{j+1}$. We decompose $H$ into $H_e=-J\sum_{j\in \text{even}} \boldsymbol{S}_j\cdot\boldsymbol{S}_{j+1}$ and $H_o=-J\sum_{j\in \text{odd}} \boldsymbol{S}_j\cdot\boldsymbol{S}_{j+1}$, where the summation is performed over either even or odd sites. Note, we do not use Trotter decomposition for $H_1$ and $H_1^\prime$ in the system discussed in Sec.~\ref{SM.3b}. There, the Hamiltonian contains three-body interactions for which one can not easily integrate the EOM analytically over a certain time window.

The EOMs generated by the two Hamiltonians can be analytically integrated over the driving duration $t$, leading to the update maps
\begin{equation}
    \tau_{e} (\Vec{S}_{j(j+1)})=\frac{1}{k_2^2}\left(
    \begin{array}{cccc}
       \relax[B_1^2+(B_2^2+B_3^2)\cos{(J_2^jk_2t)}]S_{j(j+1)}^x\\+[B_1B_2(1-\cos{(J_2^jk_2t)})+k_2B_3\sin{(J_2^jk_2t)}]S_{j(j+1)}^y\\+[B_1B_3(1-\cos{(J_2^jk_2t)})-k_2B_2\sin{(J_2^jk_2t)}]S_{j(j+1)}^z\\
       \\
    \relax[B_1B_2(1-\cos{(J_2^jk_2t)})-k_2B_3\sin{(J_2^jk_2t)}]S_{j(j+1)}^x\\+[B_2^2+(B_1^2+B_3^2)\cos{(J_2^jk_2t)}]S_{j(j+1)}^y\\+[B_2B_3(1-\cos{(J_2^jk_2t)})+k_2B_1\sin{(J_2^jk_2t)}]S_{j(j+1)}^z\\
    \\
    \relax[B_1B_3(1-\cos{(J_2^jk_2t)})+k_2B_2\sin{(J_2^jk_2t)}]S_{j(j+1)}^x\\+[B_2B_3(1-\cos{(J_2^jk_2t)})-k_2B_1\sin{(J_2^jk_2t)}]S_{j(j+1)}^y\\+[B_3^2+(B_1^2+B_2^2)\cos{(J_2^jk_2t)}]S_{j(j+1)}^z\\
    \end{array}
    \right),
\end{equation}
for $H_{e}$, $j\in even$, and
\begin{equation}
    \tau_{o} (\Vec{S}_{j(j+1)})=\frac{1}{k_2^2}\left(
    \begin{array}{cccc}
       \relax[B_1^2+(B_2^2+B_3^2)\cos{(J_2^jk_2t)}]S_{j(j+1)}^x\\+[B_1B_2(1-\cos{(J_2^jk_2t)})+k_2B_3\sin{(J_2^jk_2t)}]S_{j(j+1)}^y\\+[B_1B_3(1-\cos{(J_2^jk_2t)})-k_2B_2\sin{(J_2^jk_2t)}]S_{j(j+1)}^z\\
       \\
    \relax[B_1B_2(1-\cos{(J_2^jk_2t)})-k_2B_3\sin{(J_2^jk_2t})]S_{j(j+1)}^x\\+[B_2^2+(B_1^2+B_3^2)\cos{(J_2^jk_2t)}]S_{j(j+1)}^y\\+[B_2B_3(1-\cos{(J_2^jk_2t)})+k_2B_1\sin{(J_2^jk_2t)}]S_{j(j+1)}^z\\
    \\
    \relax[B_1B_3(1-\cos{(J_2^jk_2t)})+k_2B_2\sin{(J_2^jk_2t)}]S_{j(j+1)}^x\\+[B_2B_3(1-\cos{(J_2^jk_2t)})-k_2B_1\sin{(J_2^jk_2t)}]S_{j(j+1)}^y\\+[B_3^2+(B_1^2+B_2^2)\cos{(J_2^jk_2t)}]S_{j(j+1)}^z\\
    \end{array}
    \right),
\end{equation}
for $H_{o}$, $j\in odd$, where the average fields originating from the neighboring sites read $B_1=S_j^x+S_{j+1}^x$, $B_2=S_j^y+S_{j+1}^y$, $B_3=S_j^z+S_{j+1}^z$, and $k_2=B_1^2+B_2^2+B_3^2$.

We use the following protocol to generate the spin dynamics (second-order Suzuki-Trotter decomposition)
\begin{equation}
    H_{\text{Trotter}}(t)=\left\{
    \begin{aligned}
        H_e\quad &\text{for}\;t\in [0,\frac{T}{4}],\\
        H_o\quad &\text{for}\;t\in [\frac{T}{4},\frac{3}{4}T],\\
        H_e\quad &\text{for}\;t\in [\frac{3}{4}T,T].
    \end{aligned}
    \right
    .
\end{equation}
Time evolution after a full period can be effectively described by a Hamiltonian $\Tilde{H}$. One can prove that $\Tilde{H}=H+\mathcal{O}(T^2)$ using an inverse-frequency expansion, indicating that the error of this approximation is introduced at the order $\mathcal{O}(T^2)$. When we apply it to the whole protocol Eq.~\eqref{eq.protocol} to approximate the Heisenberg Hamiltonian, it does not change the form of $Q_{\mathrm{eff}}$, Eq.~\eqref{eq.effectiveHall} as well as the gap size of mNGs. Indeed, in principle, we can further suppress the simulation error to any required order by using higher-order Trotterization method, if needed.
\end{document}